\documentclass[aps,prb,reprint,superscriptaddress,floatfix,amsmath,amssymb]{revtex4-2}

\usepackage{graphicx}
\usepackage{bm}
\usepackage[colorlinks=true,allcolors=blue]{hyperref}

\newcommand{\braket}[2]{\langle #1 | #2 \rangle}

\begin{document}

\title{Comparing Hubbard parameters from linear-response theory and a Hartree-Fock-based approach}

\author{Wooil Yang}
\affiliation{Korea Institute for Advanced Study, Seoul 02455, Republic of Korea}
\affiliation{Oden Institute for Computational Engineering and Sciences, The University of Texas at Austin, Austin, Texas 78712, USA}

\author{Iurii Timrov}
\email[Corresponding author: ]{iurii.timrov@psi.ch}
\affiliation{PSI Center for Scientific Computing, Theory and Data, Paul Scherrer Institute, 5232 Villigen PSI, Switzerland}

\author{Francesco Aquilante}
\affiliation{Open MOLCAS Community, Geneva 1219, Switzerland}

\author{Young-Woo Son}
\email[Corresponding author: ]{hand@kias.re.kr}
\affiliation{Korea Institute for Advanced Study, Seoul 02455, Republic of Korea}

\begin{abstract}
\normalfont
Density-functional theory with on-site $U$ and inter-site $V$ Hubbard corrections (DFT+$U$+$V$) is widely used to predict properties of transition-metal and rare-earth compounds, but its accuracy depends critically on how these parameters are obtained. While they can be determined empirically, first-principles approaches offer better consistency, though their results can differ and a systematic comparison is lacking. Here, we compare two widely used approaches, linear-response theory (LRT) and the Hartree–Fock-based pseudohybrid functional formalism, applied to representative oxides (MnO, NiO, CoO, FeO, BaTiO$_3$, ZnO, and ZrO$_2$). For partially occupied transition-metal $d$ states, both methods yield comparable $U$ values, but they differ significantly for nearly empty or fully filled $d$ shells. For O-$2p$ states, LRT systematically predicts large $U$ values ($\sim$ 10eV), whereas the pseudohybrid approach gives system-dependent values. Even larger differences appear for the inter-site $V$: the former yields small values ($<1$~eV), while the latter gives larger values ($\sim 3$~eV) due to its dependence on charge redistribution. We further show that the pseudopotential choice, which defines Hubbard projectors, affects the parameters and their agreement. Overall, while parallels between these two methods exist, they rely on different assumptions, leading to variations in predictions of material properties.
\end{abstract}

\maketitle

\section*{INTRODUCTION}
\label{sec:intro}

Density-functional theory (DFT)~\cite{Hohenberg:1964, Kohn:1965} underpins modern first-principles simulations in physics, chemistry, and materials science. Despite its remarkable success, progress hinges on the development of increasingly accurate exchange-correlation (xc) functionals. Standard approximations, such as the local spin-density approximation (LSDA) and spin-polarized generalized-gradient approximation ($\sigma$-GGA), are invaluable for many systems, yet they perform poorly for transition-metal (TM) and rare-earth (RE) compounds, where localized $d$ and $f$ states suffer from large self-interaction errors (SIEs)~\cite{Perdew:1981, MoriSanchez:2006}. To address these limitations, several beyond-standard approaches have been introduced. Among the most prominent are Hubbard-corrected DFT (DFT+$U$~\cite{anisimov:1991, Liechtenstein:1995, Dudarev:1998} and its extension DFT+$U$+$V$~\cite{Campo:2010}), meta-GGA functionals (e.g., SCAN~\cite{Sun:2015} and its variants~\cite{Bartok:2019, Furness:2020}), and hybrid functionals (e.g., PBE0~\cite{Adamo:1999} and HSE06~\cite{Heyd:2003, Heyd:2006}). Combinations such as SCAN+$U$ have also been explored~\cite{Gautam:2018, Long:2020, Kaczkowski:2021, Artrith:2022}. Each strategy embodies a distinct philosophy: DFT+$U$ selectively corrects partially filled orbitals (typically $d$ and $f$)~\cite{Kulik:2006, Kulik:2008}; meta-GGA incorporates the kinetic-energy density and satisfies a broad set of exact constraints (17 in the case of SCAN); hybrid functionals mix a fraction of Fock exchange with (semi-)local functionals. While these methods improve predictive power, they each come with intrinsic trade-offs and limitations~\cite{Timrov:2022b}. Among them, Hubbard-corrected DFT has become especially popular for its ability to enhance accuracy at modest computational cost~\cite{Himmetoglu:2014}, provided reliable Hubbard parameters are available. These can be obtained either semi-empirically (by fitting to experimental data or higher-level calculations such as hybrids or $GW$~\cite{Hedin:1965}) or through systematic calibration. Examples include tuning $U$ to reproduce binary metal formation energies~\cite{Wang:2006, Kubaschewski:1993} or to match properties such as band gaps, magnetic moments, lattice parameters, and enthalpies~\cite{LeBacq:2004, Aykol:2014, Isaacs:2017}.

While semi-empirical fitting of Hubbard parameters can be effective, it becomes impractical in materials discovery where experimental data are unavailable or when multiple target properties must be simultaneously reproduced, sometimes requiring advanced optimization schemes such as Bayesian approaches~\cite{Yu:2020, Tavadze:2021}. A natural alternative is to compute Hubbard parameters from first principles using methods such as constrained DFT (cDFT)~\cite{Dederichs:1984, Mcmahan:1988, Gunnarsson:1989, Hybertsen:1989, Gunnarsson:1990, Pickett:1998, Solovyev:2005, Nakamura:2006, Shishkin:2016}, Hartree-Fock-based schemes~\cite{Mosey:2007, Mosey:2008, Andriotis:2010, Agapito:2015, Lee:2020, TancogneDejean:2020}, or the constrained random-phase approximation (cRPA)~\cite{Springer:1998, Kotani:2000, Aryasetiawan:2004, Aryasetiawan:2006}. However, predicted values can differ substantially among these approaches~\cite{Himmetoglu:2014, Aryasetiawan:2006,Liu:2023}, and systematic comparisons of their theoretical foundations remain scarce, although notable progress is beginning to emerge~\cite{Carta:2025}. Moreover, the dependence of $U$ on factors such as the choice of Hubbard projectors, pseudopotentials, oxidation state of TM ions, underlying exchange-correlation functional, local coordination environment, ligand species, and crystal structure further complicates a direct comparison of $U$ values obtained using different computational methods and setups~\cite{Kulik:2008, Shishkin:2016, Carta:2025, Mahajan:2021, Hubparam:2023, KirchnerHall:2021, Bastonero:2025, Zhou:2004b, Linscott:2018, Chai:2024, MacEnulty:2024, Moore:2024, Wang2016jcp}. Therefore, further work similar to Ref.~\citenum{Carta:2025} is needed to systematically compare and benchmark different first-principles approaches for computing Hubbard parameters using identical computational setups.

Here, we present a comparative study of two such first-principles approaches. The first is based on cDFT formulated within linear-response theory (LRT)~\cite{Cococcioni:2005}, which determines on-site $U$ by enforcing piecewise linearity of the total energy with respect to the occupation of localized orbitals (Hubbard manifold), thereby reducing SIEs~\cite{Himmetoglu:2011} and improving predictive accuracy. Standard LRT requires supercell calculations, but its reformulation within density-functional perturbation theory (DFPT)~\cite{Timrov:2018, Timrov:2021} replaces these with monochromatic reciprocal-space perturbations in a unit cell. This substantially reduces computational overhead and has enabled applications to increasingly complex systems~\cite{Mahajan:2022, Timrov:2023, Binci:2023, Gebreyesus:2023, Haddadi:2024, Bonfa:2024, Macke:2024, Warda:2026}. The second approach for computing $U$ is the Hartree-Fock-based Agapito, Curtarolo, and Buongiorno
Nardelli (ACBN0) functional~\cite{Agapito:2015} and its extension to inter-site $V$ dubbed as eACBN0~\cite{Lee:2020, TancogneDejean:2020}. This method derives Hubbard parameters directly within self-consistent-field (SCF) cycles from an approximate form of the renormalized Hartree-Fock interaction energy. Using projections of Kohn-Sham (KS) states onto localized orbitals, $U$ and $V$ are updated at each iteration of the SCF cycle with negligible additional cost relative to standard DFT. This method has been shown to capture electronic structures and dynamical properties with good accuracy~\cite{Yang:2021, Yang:2022}, and it holds promise for accurate evaluations of electron-phonon interactions~\cite{Jang:2023, Yang:2025}.

\section*{RESULTS}

\subsection*{Formalism: The DFT+$U$+$V$ approach.}
In DFT+$U$+$V$~\cite{Campo:2010, Himmetoglu:2014}, a corrective Hubbard energy $E_{U+V}$ is added to the standard DFT energy $E_{\mathrm{DFT}}$:
\begin{equation}
    E_{\mathrm{DFT}+U+V} = E_{\mathrm{DFT}} + E_{U+V} .
    \label{eq:Edft_plus_uv}
\end{equation}
Unlike DFT+$U$, which only accounts for on-site interactions scaled by $U$, DFT+$U$+$V$ also includes inter-site interactions between an atom and its neighbors, scaled by $V$. In the simplified rotationally-invariant formulation~\cite{Dudarev:1998}, the extended Hubbard correction reads~\cite{Timrov:2021}:
\begin{eqnarray}
    E_{U+V} & = & \frac{1}{2} \sum_I \sum_{\sigma m m'}
    U^I \left( \delta_{m m'} - n^{II \sigma}_{m m'} \right) n^{II \sigma}_{m' m} \nonumber \\
    & & - \frac{1}{2} \sum_{I} \sum_{J (J \ne I)}^{*} \sum_{\sigma m m'} V^{I J}
    n^{I J \sigma}_{m m'} n^{J I \sigma}_{m' m} \,,
    \label{eq:Edftu}
\end{eqnarray}
where $I$ and $J$ index atomic sites, $m$ and $m'$ denote magnetic quantum numbers, and $U^I$ and $V^{IJ}$ are the \textit{effective} on-site and inter-site Hubbard parameters, respectively (here and everywhere in the following, $U^I$ denotes the difference between the screened Coulomb interaction parameter and the Hund's exchange interaction parameter). The asterisk denotes that, for each atom $I$, the sum over $J$ includes all neighbors up to a defined cutoff distance. The generalized occupation matrices $n^{IJ \sigma}_{m m'}$ are obtained by projecting KS states $\psi^\sigma_{v,\mathbf{k}}$ onto localized orbitals $\phi^I_m(\mathbf{r})$:
\begin{equation}
    n^{I J \sigma}_{m m'} = \sum_{v,\mathbf{k}} f^\sigma_{v,\mathbf{k}}
    \braket{\psi^\sigma_{v,\mathbf{k}}}{\phi^{J}_{m'}} \braket{\phi^{I}_{m}}{\psi^\sigma_{v,\mathbf{k}}} \,,
    \label{eq:occ_matrix_0}
\end{equation}
where $v$ and $\sigma$ label bands and spin, $\mathbf{k}$ runs over the first Brillouin zone (BZ), and $f^\sigma_{v,\mathbf{k}}$ are the KS occupations. In Eq.~\eqref{eq:occ_matrix_0}, $m \in I$ and $m' \in J$, meaning that $m$ runs over the magnetic quantum numbers associated with a fixed principal quantum number $n$ and orbital angular momentum $l$ of atom $I$, and analogously $m'$ for atom $J$. The corresponding generalized KS equation can be written as~\cite{Campo:2010},
\begin{equation}
     \left(\hat{\mathcal H}_\mathrm{KS}^\sigma +
     \hat{\mathcal V}_{U+V}^\sigma \right)|\psi_{v,{\bf k}}^\sigma\rangle
     = \varepsilon_{v,{\bf k}}^\sigma |\psi_{v,{\bf k}}^\sigma\rangle,
\end{equation}
where $\hat{\mathcal H}_\mathrm{KS}^\sigma$ is the KS Hamiltonian of the standard DFT, $\varepsilon_{v,{\bf k}}^\sigma$ is the $v$-th band energy at momentum ${\bf k}$ and spin $\sigma$, and $\hat{\mathcal V}_{U+V}^\sigma$ is the Hubbard potential that is derived from the Hubbard energy~\eqref{eq:Edftu} and which reads~\cite{Timrov:2021}:
\begin{eqnarray} 
\hat{\mathcal V}_{U+V}^\sigma & = &
    \sum_{I}\sum_{mm'}\frac{U^I}{2}
    \left(
    \delta_{mm'}-2n^{II\sigma}_{mm'}
    \right) 
    |\phi^I_m\rangle
    \langle\phi^I_{m'}| \nonumber \\
    & & - \sum_{I} \sum_{J (J \ne I)}^{*}\sum_{mm'} V^{IJ} n^{IJ\sigma}_{mm'}
    |\phi^I_m\rangle
    \langle \phi^J_{m'}|.
    \label{eq:Vdftuv}
\end{eqnarray}
In Eqs.~\eqref{eq:Edftu} and \eqref{eq:Vdftuv}, the effective on-site $U^I$ and inter-site $V^{IJ}$ terms act in opposition when both parameters are positive: the on-site term favors electron localization on atomic sites and suppresses hybridization, whereas the inter-site term enhances hybridization with neighboring atoms. Accurate first-principles evaluation of $U^I$ and $V^{IJ}$ is therefore crucial to correctly describe the balance between localization and hybridization. 

It is important to note that the computed values of $U^I$ and $V^{IJ}$ depend on the choice of Hubbard projector functions $\phi^I_{m}(\mathbf{r})$ used to define the occupation matrix in Eq.~\eqref{eq:occ_matrix_0} and the Hubbard potential in Eq.~\eqref{eq:Vdftuv}~\cite{Lee:2020, KirchnerHall:2021, Mahajan:2021}. In the following, we briefly describe how Hubbard parameters are defined and computed within LRT and eACBN0. For both methods, we employ the same projector functions, namely atomic orbitals orthogonalized using the L\"owdin method~\cite{Lowdin:1950, Mayer:2002}, to ensure a fair and consistent comparison of the computed $U^I$ and $V^{IJ}$ values.

%
Within the LRT framework, the effective Hubbard parameters are defined as~\cite{Cococcioni:2005, Campo:2010, Bastonero:2025}:
\begin{equation}
U^I = \left(\chi_0^{-1} - \chi^{-1}\right)_{II} \,,
\label{eq:Ucalc}
\end{equation}
and
\begin{equation}
V^{IJ} = \left(\chi_0^{-1} - \chi^{-1}\right)_{IJ} \,,
\label{eq:Vcalc}
\end{equation}
where $\chi_0$ and $\chi$ are the bare and self-consistent response matrices, describing how the occupation matrices respond to localized electronic perturbations. The self-consistent response matrix is given by
\begin{equation}
\chi_{IJ} = \sum_{m\sigma} \frac{dn^{I \sigma}_{mm}}{d\alpha^J} \,,
\end{equation}
where $n^{I\sigma}_{mm'} \equiv n^{II\sigma}_{mm'}$ denotes the on-site occupation matrix, and $\alpha^J$ is the strength of the perturbation on the $J$th site. The bare response $\chi_0$ is defined analogously but evaluated before the self-consistent adjustment of the Hartree and xc potentials.

DFPT allows this calculation to be performed without supercells, by reformulating LRT in terms of monochromatic reciprocal-space perturbations in the unit cell. This significantly reduces computational cost~\cite{Timrov:2018, Timrov:2021}. The response of the occupation matrix is then computed as:
\begin{equation}
\frac{dn^{I \sigma}_{mm'}}{d\alpha^J} = \frac{1}{N_{\mathbf{q}}}\sum_{\mathbf{q}}^{N_{\mathbf{q}}} e^{i\mathbf{q}\cdot(\mathbf{R}_{l} - \mathbf{R}_{l'})}\Delta_{\mathbf{q}}^{s'} n^{s \sigma}_{mm'} \,,
\label{eq:dnq}
\end{equation}
where $\mathbf{q}$ is the wavevector of the monochromatic perturbation, $N_\mathbf{q}$ is the total number of perturbations, and $\Delta_{\mathbf{q}}^{s'} n^{s \sigma}_{mm'}$ is the lattice-periodic response of the occupation matrices to a $\mathbf{q}$-specific perturbation. Here, $I\equiv(l,s)$ and $J\equiv(l',s')$, with $s$ and $s'$ indexing atoms within the unit cell and $l$ and $l'$ indexing unit cells; $\mathbf{R}_l$ and $\mathbf{R}_{l'}$ are the corresponding Bravais lattice vectors. The quantities $\Delta{\mathbf{q}}^{s'} n^{s \sigma}_{mm'}$ are obtained from the response KS wavefunctions, which are computed by solving $\mathbf{q}$-specific Sternheimer equations.
Further details on the DFPT implementation can be found in Refs.~\citenum{Timrov:2018, Timrov:2021}. The efficiency of DFPT for calculating $U^I$ and $V^{IJ}$ has been demonstrated in numerous studies, with several recent highlights~\cite{Gelin:2024, Uhrin:2025, Bastonero:2025, Chang:2025, Binci:2025, dosSantos:2026}.

%
In the eACBN0 method, the effective Hubbard parameters are defined by equating the corrective Hubbard energy [Eq.~\eqref{eq:Edftu}] with the Hartree-Fock interaction energy~\cite{Agapito:2015, Lee:2020}. This leads to the following expressions:
\begin{eqnarray}
    U^{I} & = & 
    \frac{\sum\limits_{\{m\},\sigma\sigma'} \bar{P}^{II,\sigma}_{mm'}\bar{P}^{II,\sigma'}_{m''m'''}(mm'|m''m''')}{\sum\limits_{mm',\sigma}{n^{II,\sigma}_{mm}n^{II,-\sigma}_{m'm'}}+\sum\limits_{m \neq m',\sigma}{n^{II,\sigma}_{mm}n^{II,\sigma}_{m'm'}}}, \nonumber \\
    & - & \frac{\sum_{\{m\},\sigma} \bar{P}^{II,\sigma}_{mm'}\bar{P}^{II,\sigma}_{m''m'''}(mm'''|m''m')}{\sum_{m \neq m',\sigma}{n^{II,\sigma}_{mm}n^{II,\sigma}_{m'm'}} } \,,
    \label{eq:U_acbn0}
\end{eqnarray}
and
\begin{eqnarray}
        V^{IJ} & = & \frac{1}{2}\frac{\sum\limits_{\substack{mm',\sigma\sigma'}} [\bar{P}^{II,\sigma}_{mm}\bar{P}^{JJ,\sigma'}_{m'm'}- \delta_{\sigma\sigma'}\bar{P}^{IJ,\sigma}_{mm'}\bar{P}^{JI,\sigma'}_{m'm}]}{\sum\limits_{\substack{mm', \sigma\sigma'}}({n^{II,\sigma}_{mm}n^{JJ,\sigma'}_{m'm'}}-\delta_{\sigma\sigma'}{n^{IJ,\sigma}_{mm'}n^{JI,\sigma'}_{m'm}})} \nonumber   
    \\
    & &\times (mm|m'm') \,,
    \label{eq:V_acbn0}
\end{eqnarray}
where $(mm'|m''m''')$ represents the bare Coulomb interaction between electrons with $m, m'$ of atom $I$ and $m'', m'''$ of atom $J$. 
The renormalized density matrix $\bar{P}^{IJ,\sigma}_{mm'}$ is defined as
\begin{equation}
\bar{P}^{IJ,\sigma}_{mm'} = \sum_{v,\mathbf{k}} f^{\sigma}_{v,\mathbf{k}} \, \bar{N}^{IJ,\sigma}_{v,\mathbf{k}} \langle \psi^{\sigma}_{v,\mathbf{k}} | \phi^{J}_{m'} \rangle \langle \phi^{I}_{m} | \psi^{\sigma}_{v,\mathbf{k}} \rangle \,,
\label{eq:rdm}
\end{equation}
where $\bar{N}^{IJ,\sigma}_{v,\mathbf{k}}$ is the renormalized occupation number defined as
\begin{eqnarray}
\bar{N}^{IJ,\sigma}_{v,\mathbf{k}} & = & \sum_{\{I\}}\sum_{m} \langle \psi^{\sigma}_{v,\mathbf{k}} |  \phi^{I}_{m} \rangle \langle  \phi^{I}_{m} | \psi^{\sigma}_{v,\mathbf{k}} \rangle  \nonumber \\
& & + \sum_{\{J\}}\sum_{m'} \langle \psi^{\sigma}_{v,\mathbf{k}} |  \phi^{J}_{m'} \rangle \langle  \phi^{J}_{m'} | \psi^{\sigma}_{v,\mathbf{k}} \rangle \,, 
\label{eq:rom}
\end{eqnarray}
where $m \in I$ and $m' \in J$, and the summations $\sum_{\{I\}}$ and $\sum_{\{J\}}$ mean that we sum up contributions coming from all equivalent atoms of the same type. For the case of $U^I$-only calculations ($I=J$), the second term in Eq.~\eqref{eq:rom} is dropped~\cite{Lee:2020}, so $\bar{N}^{II,\sigma}_{v,\mathbf{k}}$ reduces to the Mulliken charge of atom $I$ in state $\psi^{\sigma}_{v,\mathbf{k}}(\mathbf{r})$. Moreover, $\bar{N}^{IJ,\sigma}_{v,\mathbf{k}}$ can be viewed as a position-dependent mixing parameter within the Hartree-Fock approximation, encoding the effective on-site and inter-site interactions in the localized subspace. It is important to note that for a fixed ground-state density, the $U^I$ values computed in ACBN0 remain unchanged upon inclusion of inter-site $V^{IJ}$ terms in eACBN0. However, in self-consistent implementations of eACBN0, a coupling between $U^I$ and $V^{IJ}$ emerges because the ground state is obtained under the influence of both on-site and inter-site terms in the Hubbard potential~\eqref{eq:Vdftuv}, with self-consistency enforced for both the electron density and the Hubbard parameters. Further details on the eACBN0 formalism can be found in Refs.~\citenum{Agapito:2015,Lee:2020, TancogneDejean:2020}.

Finally, we comment on the normalization entering Eq.~\eqref{eq:rom}. Upon enlarging the unit cell, the number of equivalent atomic orbital projectors included in the sums increases. However, each KS wavefunction is normalized to unity within the unit cell, and Eq.~\eqref{eq:rom} is defined at the level of individual KS states (there is no summation over $v$ and $\mathbf{k}$). As a result, when the cell is enlarged, each KS state delocalizes over a larger number of equivalent atoms, reducing its projection on each atom, while the sum over all equivalent atoms increases proportionally. These two effects exactly compensate, so that the renormalized occupation number in Eq.~\eqref{eq:rom} remains unchanged for the corresponding KS states. Consequently, the Hubbard parameters obtained from Eqs.~\eqref{eq:U_acbn0} and~\eqref{eq:V_acbn0} are invariant with respect to the choice of unit cell.

\subsection*{Overview of computational protocols}

In this section, we present results for the computed on-site $U$ and inter-site $V$ Hubbard parameters across three classes of systems:
$i)$ TM oxides with partially occupied $d$ states (MnO, FeO, CoO, and NiO);
$ii)$ compounds with nominally empty $d$ states (ZrO$_2$ and BaTiO$_3$); and
$iii)$ a compound with fully occupied $d$ states (ZnO).
These categories allow us to assess similarities and differences between LRT and eACBN0 under distinct electronic configurations. 

\begin{figure}[t!]
\centering
\includegraphics[width=1\columnwidth]{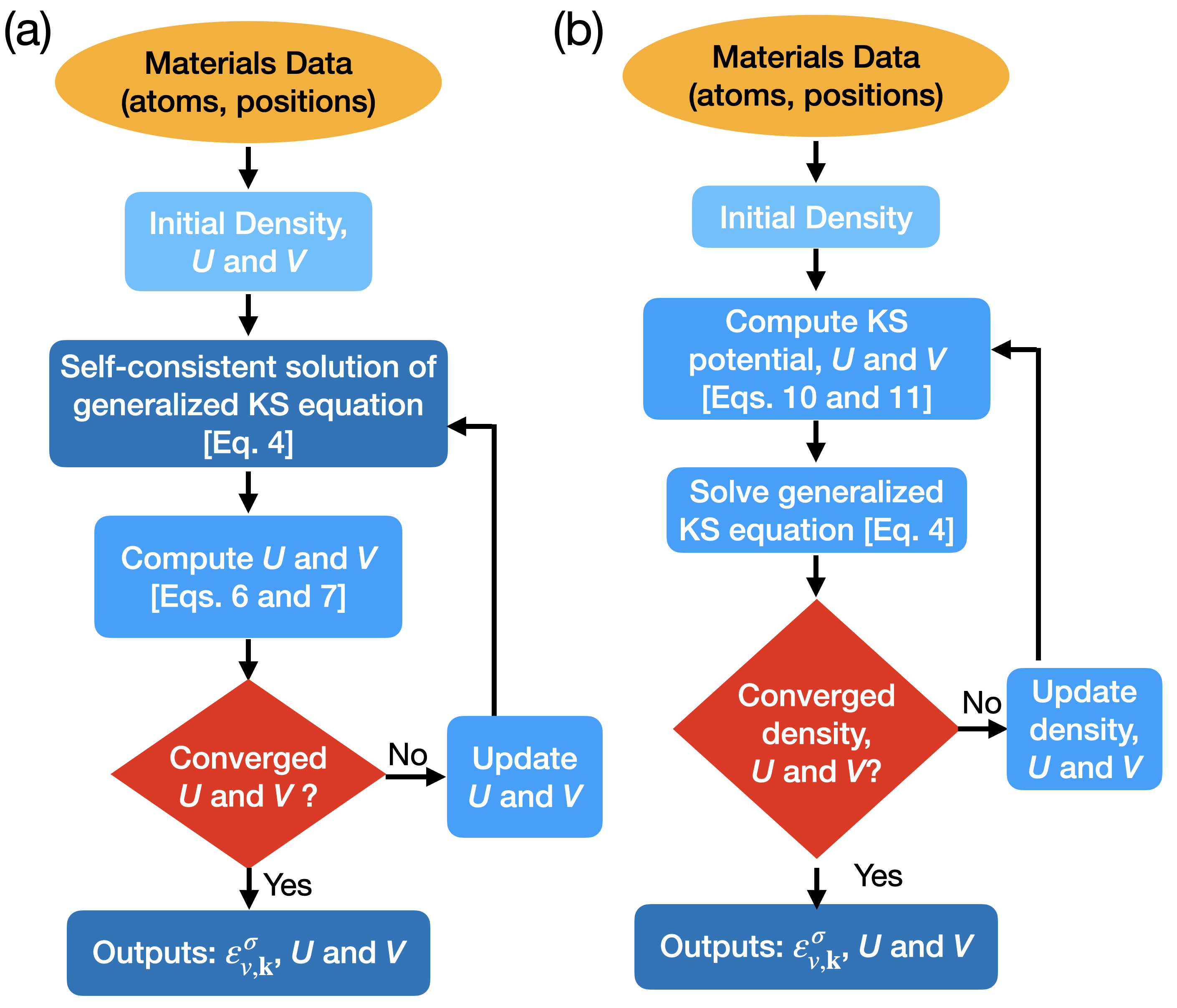}
\caption{Protocol for computing Hubbard parameters within the ``self-consistent'' scheme using (a) LRT (DFPT)~\cite{Timrov:2021} and (b) eACBN0~\cite{Lee:2020}.
In the LRT workflow, Hubbard parameters are evaluated self-consistently after the electronic charge density has been fully converged. In contrast, the eACBN0 scheme updates the charge density and the Hubbard parameters simultaneously until convergence is reached. In both workflows, atomic positions are fixed and no structural relaxations are performed.}
\label{Fig:flowchart}
\end{figure}

Hubbard parameters can be computed at different levels of self-consistency. The simplest approach is a ``one-shot'' calculation, where the parameters are evaluated using the converged DFT charge density while keeping the geometry fixed. 
A more advanced ``self-consistent'' approach iteratively updates both the charge density and the corresponding $U$ and $V$ values, but still at a fixed geometry. Finally, in the ``structurally self-consistent'' scheme, structural relaxations (atomic positions and unit-cell parameters) are included in the self-consistent cycle~\cite{Timrov:2021, Bastonero:2025}.
In the main text, we report results obtained with the ``self-consistent'' scheme, i.e.\ calculations in which the electronic degrees of freedom are converged together with the Hubbard parameters while the geometry remains fixed. 
In this approach, all $U$ and $V$ values are determined consistently with the self-consistent charge density~\cite{Timrov:2021, Lee:2020}, as illustrated in Fig.~\ref{Fig:flowchart}. 
This allows us to isolate purely electronic effects from structural ones and to assess the influence of extended Hubbard interactions in LRT and eACBN0 with high internal consistency. For completeness, in the Supplemental Information (SI) we also present results obtained using the ``structurally self-consistent'' scheme, in which the self-consistent cycle includes full structural optimization with updated Hubbard parameters and charge density (see Sec.~S1 and S2 in the SI).  In this case, the crystal structure, the Hubbard parameters, and the charge density are iterated to convergence, yielding mutually consistent electronic and crystal structures. All computational details are provided in the Methods section. 

We note that results obtained using the ``one-shot'' scheme are not reported in this work. In this scheme, the occupation matrices of Eq.~\eqref{eq:occ_matrix_0} and Kohn-Sham wavefunctions employed to compute the Hubbard parameters (via Eqs.~\eqref{eq:Ucalc} and \eqref{eq:Vcalc} for LRT and Eqs.~\eqref{eq:U_acbn0} and \eqref{eq:V_acbn0} for eACBN0) differ from those that are obtained during the final DFT+$U$+$V$ ground-state SCF calculation, which is determined by the Hubbard energy [Eq.~\eqref{eq:Edftu}] and the Hubbard potential [Eq.~\eqref{eq:Vdftuv}]. This difference between the Hubbard parameters evaluation step and the subsequent materials-property calculations makes the ``one-shot'' scheme not consistent. Moreover, the eACBN0 method is never used in a ``one-shot'' form in practice. To avoid potential ambiguity for the reader, we therefore do not include ``one-shot'' results in this work.

Using the computed self-consistent Hubbard parameters, we perform DFT+$U$ and DFT+$U$+$V$ calculations to evaluate various material properties. The band gap serves as our primary benchmark for accuracy. The experimental band gaps used for comparison were taken from the literature and were obtained using different techniques depending on the material: X-ray emission spectroscopy (XES) and X-ray absorption spectroscopy (XAS) for ZnO~\cite{Dong2004prb}, MnO, CoO, and NiO~\cite{Kurmaev2008prb}; Photoemission Spectroscopy (PES) and Bremsstrahlung isochromat spectroscopy (BIS) for MnO~\cite{Elp15301991prb}, CoO~\cite{Elp60901991prb}, and NiO~\cite{Sawatzky1984prl}; Conductivity for MnO~\cite{Drabkin:1968}, CoO~\cite{Gvishi:1972}, and NiO~\cite{ksendzov:1965}; Ultraviolet photoelectron spectroscopy (UPS) and inverse photoemission spectroscopy (IPS) for ZrO$_2$~\cite{Bersch:2008}; and optical absorption edge measurements for FeO~\cite{Bowen1975JSSC} and BaTiO$_3$~\cite{Wemple:1970}. We assume that excitonic effects are weak in BaTiO$_3$ due to the relatively large dielectric screening, so that the measured absorption edge closely approximates the fundamental band gap with minimal influence from excitonic states. In contrast, FeO exhibits substantially weaker dielectric screening than BaTiO$_3$, which in principle could lead to more pronounced electron-hole interactions. However, we make an assumption that the optical response in FeO is dominated by the valence-conduction interband excitations, so that well-defined excitonic features are not expected to be the primary determinant of the absorption edge. In the absence of reliable experimental determinations of the fundamental gap in BaTiO$_3$ and FeO, we therefore use the reported optical absorption edge as an approximate proxy, while noting that it may include additional many-body effects. For magnetic oxides we also examine magnetic moments, while for BaTiO$_3$ we additionally analyze structural properties relevant to ferroelectricity. The results are systematically compared with experiments.

Local and semilocal xc functionals (LSDA, $\sigma$-GGA) notoriously underestimate band gaps in transition-metal oxides (TMOs) due to strong SIEs, particularly on TM ions. DFT+$U$ can alleviate this issue, but only when the correction is applied to states that define the gap edges (for insulators) or to states near the Fermi level (for spurious DFT metals)~\cite{KirchnerHall:2021}. The inter-site $V$ correction can further improve band gaps in covalent systems, but again only when applied to orbitals directly involved in gap formation. Prior work supports these points: Ref.~\citenum{KirchnerHall:2021} showed that DFT+$U$ with LRT-derived $U$ improves agreement with experimental band gaps compared to (semi-)local functionals, while DFT+$U$+$V$ with LRT-derived parameters yields even better accuracy in TMOs with strong covalency~\cite{Mahajan:2021, Mahajan:2022, Timrov:2023, Gebreyesus:2023}. For $p$-block semiconductors without TM ions, Ref.~\citenum{Lee:2020} demonstrated that DFT+$U$+$V$ with eACBN0-derived parameters accurately reproduces experimental band gaps. These insights provide the foundation for our analysis below.

\subsection*{MnO, FeO, CoO, and NiO}

TMOs are well-known systems where SIEs are prominently manifested due to localized $3d$ orbitals. Standard DFT calculations at the LDA or GGA level typically exhibit considerable discrepancies compared to experimental measurements, motivating the use of the DFT+$U$ or DFT+$U$+$V$ method. Particularly, $3d$ electrons of TM ions exhibit significant spatial localization, which can be effectively treated using the Hubbard $U$ corrections, while $V$ might be necessary to describe their hybridization with ligands. Therefore, we select representative $3d$ TMOs (MnO, FeO, CoO, and NiO) as first targets and apply two Hubbard parameter calculation methods, LRT and eACBN0, to understand the trends in Hubbard parameters and their influence on materials properties such as band gaps and magnetic moments. 

\begin{figure*}[t!]
\centering
\includegraphics[width=0.8\textwidth]{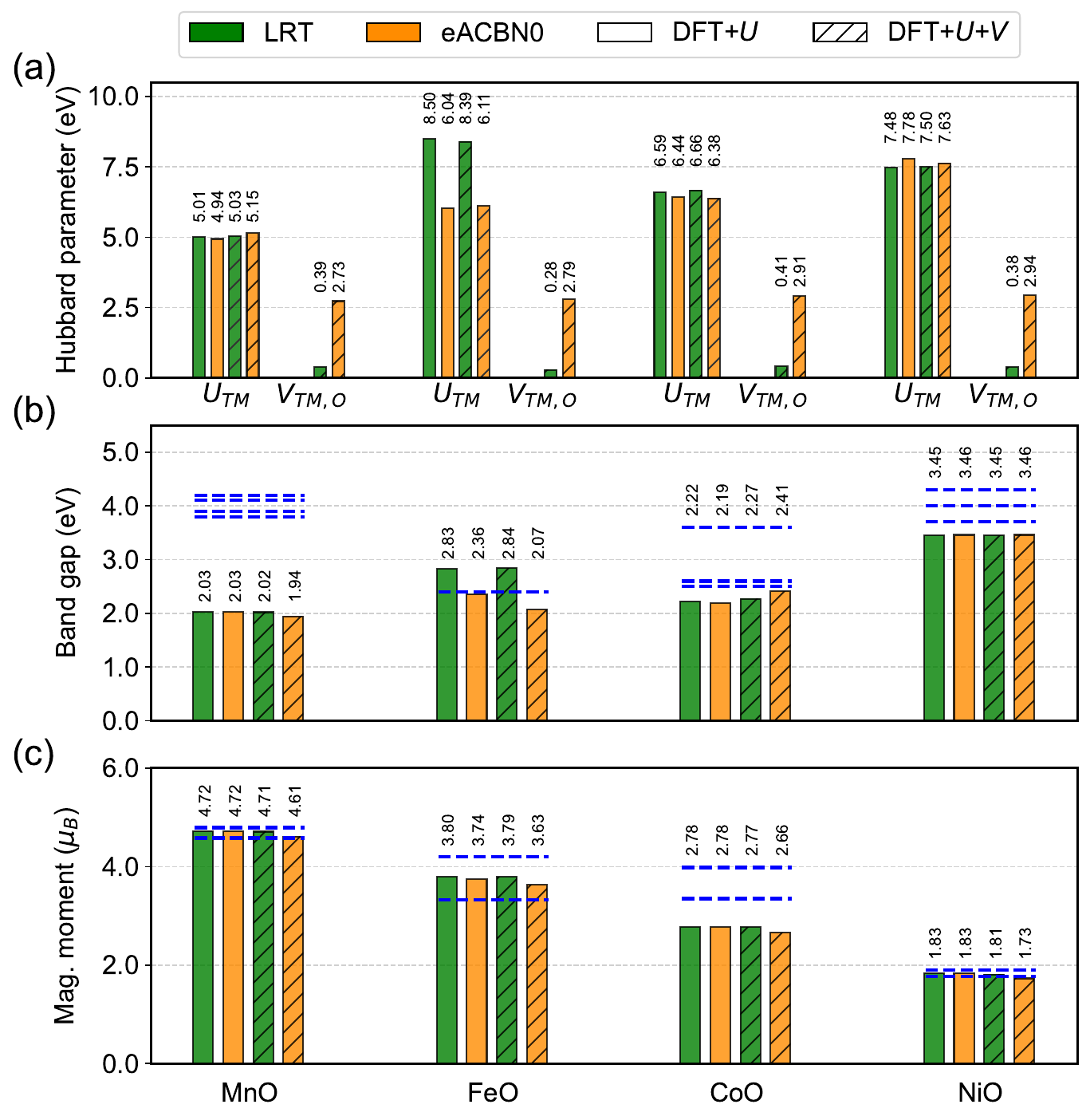}
\caption{DFT+$U$ and DFT+$U$+$V$ self-consistent calculations for TMOs without on-site corrections on O-$2p$ states ($U_{\mathrm{O}}$), using the workflow described in Fig.~\ref{Fig:flowchart}. 
(a) On-site Hubbard $U_{\mathrm{TM}}$ parameter for $3d$ orbitals of TM ions and inter-site Hubbard $V_{\mathrm{TM,O}}$ parameter between $3d$ orbitals of TM ions and $2p$ orbitals of oxygen, (b) band gap, and (c) magnetic moment. LRT and eACBN0 denote the method used to obtain the Hubbard parameters. The experimental band gaps and magnetic moments for MnO~\cite{Kurmaev2008prb,Elp15301991prb,Drabkin:1968,Cheetham1983prb,Fender1968JCP}, FeO~\cite{Bowen1975JSSC,Roth1958prb,Battle1979JPCSSP}, CoO~\cite{Kurmaev2008prb,Elp60901991prb,Gvishi:1972,Khan1970prb,Jauch2001prb}, and NiO~\cite{Kurmaev2008prb,Sawatzky1984prl,ksendzov:1965,Cheetham1983prb,Fender1968JCP} are indicated by the dashed blue lines in panels (b) and (c), respectively.}
\label{Fig:TMO_wo_O}
\end{figure*}

\begin{figure*}[t!]
\centering
\includegraphics[width=0.8\textwidth]{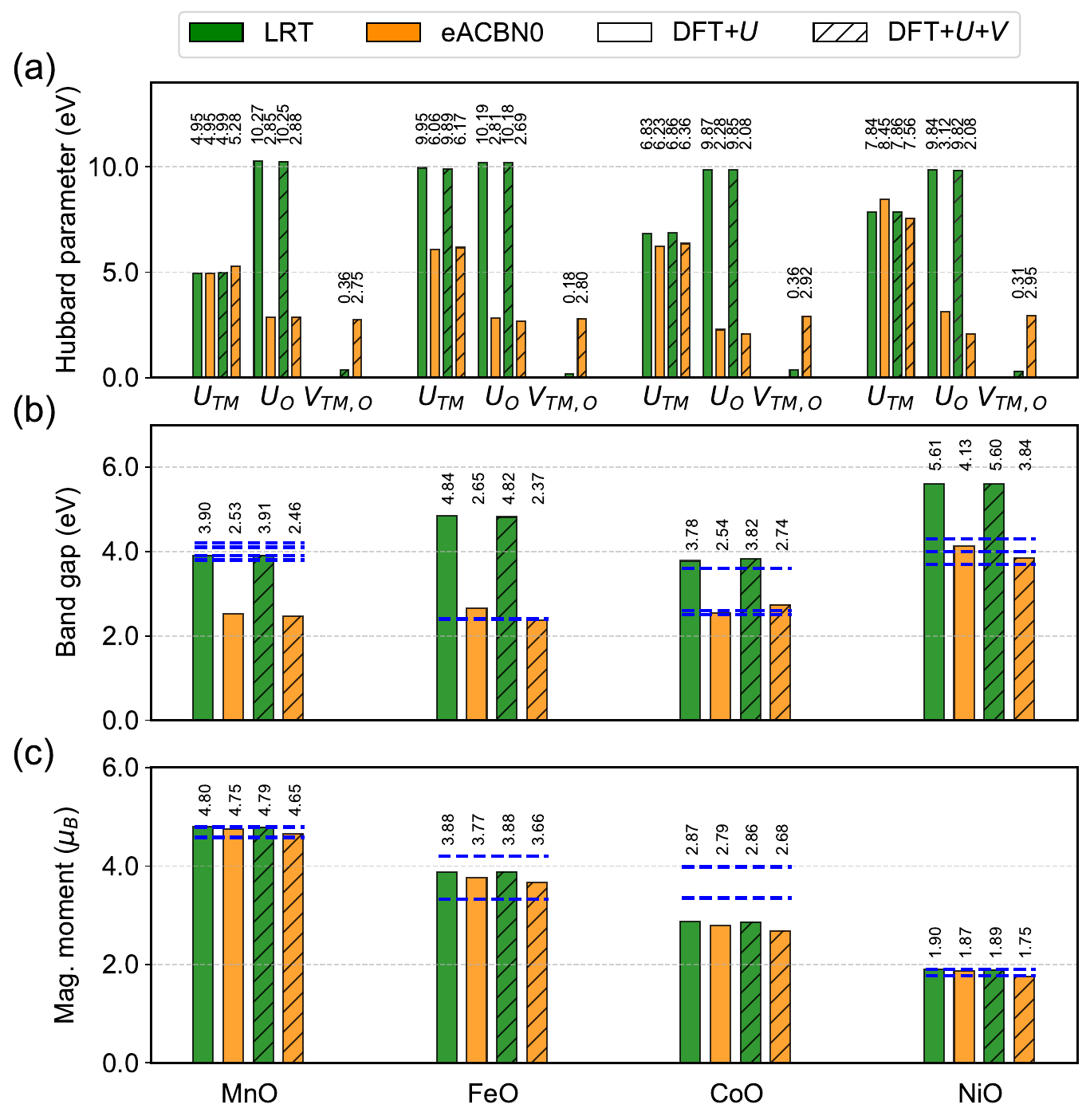}
\caption{DFT+$U$ and DFT+$U$+$V$ self-consistent calculations for TMOs with on-site $U_{\mathrm{O}}$ corrections on O-$2p$ states. (a) On-site Hubbard $U_{\mathrm{TM}}$ and $U_{\mathrm{O}}$ parameters for $3d$ orbitals of TM ions and $2p$ orbitals of oxygen, respectively, and inter-site Hubbard $V_{\mathrm{TM,O}}$ parameter between $3d$ orbitals of TM ions and $2p$ orbitals of oxygen, (b) band gap, and (c) magnetic moment. LRT and eACBN0 denote the method used to obtain the Hubbard parameters. The experimental band gaps and magnetic moments for MnO~\cite{Kurmaev2008prb,Elp15301991prb,Drabkin:1968,Cheetham1983prb,Fender1968JCP}, FeO~\cite{Bowen1975JSSC,Roth1958prb,Battle1979JPCSSP}, CoO~\cite{Kurmaev2008prb,Elp60901991prb,Gvishi:1972,Khan1970prb,Jauch2001prb}, and NiO~\cite{Kurmaev2008prb,Sawatzky1984prl,ksendzov:1965,Cheetham1983prb,Fender1968JCP} are indicated by the dashed blue lines in panels (b) and (c), respectively.}
\label{Fig:TMO_with_O}
\end{figure*}

To compare the two methods under as equivalent conditions as possible, we perform LRT and eACBN0 calculations using the self-consistent scheme described in Fig.~\ref{Fig:flowchart}. The resulting Hubbard parameters are reported in Figs.~\ref{Fig:TMO_wo_O} and \ref{Fig:TMO_with_O}. We first focus on the $U$ values obtained within DFT+$U$, shown as unhatched bar plots in Fig.~\ref{Fig:TMO_wo_O}.
We find that the $U_{\mathrm{TM}}$ values obtained from LRT and ACBN0 are in excellent agreement for MnO, CoO, and NiO, with differences not exceeding 0.3~eV. In contrast, a significantly larger discrepancy of about 1.5~eV is observed for FeO. To understand this behavior, it is important to consider the role of the pseudopotentials used in our calculations. Specifically, we employ pseudopotentials from the GBRV library (v1.5)~\cite{Garrity:2014}, where the TM ions are represented in ionized configurations: $3d^5 4s^{1.5}$ for Mn, $3d^5 4s^{2}$ for Fe, $3d^7 4s^{1}$ for Co, and $3d^8 4s^{0}$ for Ni. Notably, for Mn, Co, and Ni, the ionization affects the $4s$ shell relative to the neutral atom, whereas for Fe it affects the $3d$ shell. This distinction is crucial because the computed $U_{\mathrm{TM}}$ values are sensitive to the choice of pseudopotentials when Hubbard projectors are constructed from (orthogonalized) atomic orbitals derived from them, as is the case here. In particular, previous studies have shown that, within LRT, variations in the ionization of the $3d$ states can lead to changes in $U_{\mathrm{TM}}$ of up to 2~eV for Fe~\cite{Kulik:2008}. Therefore, to further investigate this effect, in Sec.~S3 in the SI we report results obtained using the PSlibrary (v0.3.1)~\cite{DalCorso:2014}, in which the TM pseudopotentials are not ionized. In this case, the agreement between LRT and ACBN0 for FeO improves significantly, with deviations reduced to about 0.3~eV. However, this improvement comes at the expense of a larger discrepancy for MnO, CoO, and NiO, reaching up to 1.8~eV for NiO. These observations highlight the importance of paying attention to pseudopotentials when comparing Hubbard parameters obtained from different methods. Overall, $U_{\mathrm{TM}}$ computed within LRT is considerably more sensitive to the choice of pseudopotentials (and thus to the definition of Hubbard projectors), whereas ACBN0 exhibits a much weaker dependence, with variations not exceeding 0.4~eV for the TMOs considered here. Returning to Fig.~\ref{Fig:TMO_wo_O}, and excluding the LRT result for FeO, a clear trend emerges: $U_{\mathrm{TM}}$ increases from MnO to NiO, following the increasing occupation of the $3d$ shell. A notable feature of the eACBN0 results is that the converged $U_{\mathrm{TM}}$ values closely track the nominal electron count of the divalent TM cations. This behavior is consistent with the eACBN0 formalism, in which the effective on-site interaction is constructed from renormalized occupations. Consequently, $U_{\mathrm{TM}}$ naturally reflects the filling of the TM-$3d$ states.

We now discuss the magnetic moments and band gaps obtained within the DFT+$U$ framework, shown in Figs.~\ref{Fig:TMO_wo_O}(b) and (c). Starting with the magnetic moments, we observe a decrease with increasing $3d$ shell occupancy, i.e., when going from MnO to NiO, as expected. Since the values of $U_{\mathrm{TM}}$ obtained from LRT and ACBN0 are very similar for these systems (with the exception of FeO, as discussed above), the corresponding magnetic moments are also in close agreement. Notably, even for FeO, where the difference in $U_{\mathrm{TM}}$ between LRT and ACBN0 is relatively large, the computed magnetic moments remain very similar, indicating that they are not strongly sensitive to the value of $U_{\mathrm{TM}}$. Overall, the calculated magnetic moments are in good agreement with available experimental data, with the largest discrepancy observed for CoO, of about $0.6 - 1.2$~$\mu_\mathrm{B}$. Turning to the band gaps, we find that the predictions based on $U_{\mathrm{TM}}$ from LRT and ACBN0 are generally consistent, with the largest deviation again occurring for FeO, because it shows the largest difference in $U_{\mathrm{TM}}$. The computed band gaps are in relatively good agreement with some experimental data for FeO, CoO, and NiO, while larger discrepancies are found for MnO amounting to approximately 2~eV. Focusing on MnO, the lowest conduction band is dominated by Mn-$4s$ character and is therefore largely insensitive to the Hubbard $U_{\mathrm{TM}}$ applied to Mn-$3d$ states. The primary effect of $U_{\mathrm{TM}}$ is to increase the splitting between occupied and unoccupied Mn-$3d$ states. This is confirmed by the partial density of states (PDOS), which shows that, with increasing $U_{\mathrm{TM}}$, the occupied $3d$ states shift to lower energies, while the unoccupied $3d$ states (located above the conduction band minimum) shift to higher energies~\cite{Yang:2022}. This mechanism is general and also applies to other TM oxides such as FeO, CoO, and NiO, which exhibit a significant TM-$4s$ contribution at the conduction band minimum at the $\Gamma$ point. For these materials, although the energy splitting between occupied and unoccupied $3d$ states increases proportionally with $U_{\mathrm{TM}}$, the resulting change in the fundamental band gap remains significantly smaller.

We now turn to the case where inter-site Hubbard interactions ($V_{\mathrm{TM,O}}$) between TM-$3d$ and O-$2p$ states are included. Within LRT, the determination of $V_{\mathrm{TM,O}}$ requires oxygen to be treated as an additional perturbed subspace. As a result, the response matrices are modified, since additional screening channels involving O states are activated. In LRT, $U_{\mathrm{TM}}$ and $U_{\mathrm{O}}$ appear as diagonal elements, and $V_{\mathrm{TM,O}}$ as off-diagonal elements of the matrix obtained by inverting the interacting and non-interacting response matrices, $\chi$ and $\chi_{0}$ [Eqs.~\eqref{eq:Ucalc} and \eqref{eq:Vcalc}]. Once both TM and O sites are perturbed, the structure of the response matrices in DFT+$U$ and DFT+$U$+$V$ becomes effectively similar: both schemes capture the same screening pathways involving TM and O states. An important practical aspect is whether the on-site Hubbard term $U_{\mathrm{O}}$ should be retained in the self-consistent DFT+$U$ and DFT+$U$+$V$ cycles or excluded. In Fig.~\ref{Fig:TMO_wo_O} we present results obtained when $U_{\mathrm{O}}$ is disregarded, while Fig.~\ref{Fig:TMO_with_O} shows the corresponding results when $U_{\mathrm{O}}$ is included. The choice of whether to include $U_{\mathrm{O}}$ is not unique in practical calculations. It depends on the material under study, the property of interest, and the specific method used to compute the Hubbard parameters~\cite{Lee:2020, TancogneDejean:2020, KirchnerHall:2021, Orhan:2020}.

When the inter-site Hubbard interactions ($V_{\mathrm{TM,O}}$) between TM-$3d$ and O-$2p$ states are included, the differences between the two methods remain almost unchanged. In Fig.~\ref{Fig:TMO_wo_O}, the results with the $V_{\mathrm{TM,O}}$ term are illustrated with hatched patterns. 
We observe that the inclusion of inter-site interactions within both LRT and ACBN0 does not significantly modify the values of $U_{\mathrm{TM}}$, as evidenced by the close agreement between the hatched and unhatched bars of the same color in Fig.~\ref{Fig:TMO_wo_O}(a). This weak sensitivity originates from the self-consistent determination of the Hubbard parameters: although the inclusion of $V$ can modify the $U_{\mathrm{TM}}$ values at the initial stage of the calculation, these differences are largely reduced once the Hubbard parameters and the charge density are updated self-consistently. We note, however, that this behavior depends on system. In other materials, the inclusion of inter-site interactions may lead to more pronounced changes in $U_{\mathrm{TM}}$ even within a self-consistent scheme (see e.g. Fig.~\ref{Fig:ZrO2_wo_O}).
Next, LRT yields relatively small $V_{\mathrm{TM,O}}$ values (below 1~eV), resulting in negligible differences in band gaps and magnetic moments whether $V_{\mathrm{TM,O}}$ term is included or not in the total energy calculation. In contrast, eACBN0 consistently produces larger $V_{\mathrm{TM,O}}$ values (close to 3~eV) across all TMOs, resulting in more noticeable changes in band gaps compared to LRT, while the magnetic moments change by a small margin. This effect arises because the $V_{\mathrm{TM,O}}$ term facilitates electron delocalization into the overlapping regions between atomic pairs. This counteracts the strong localization favored by the $U_{\mathrm{TM}}$ term and thus reduces the local magnetic moment, as previously shown through charge density difference analysis in our earlier studies~\cite{Yang:2022}.

Figure~\ref{Fig:TMO_with_O} extends our analysis by including a Hubbard correction on the O-$2p$ states ($U_{\mathrm{O}}$), computed with both LRT and eACBN0. Within LRT, the resulting $U_{\mathrm{O}}$ values are large ($\sim 10$~eV), consistently with previous studies~\cite{KirchnerHall:2021}. This can be understood in terms of oxygen’s large chemical hardness, arising from its high ionization energy and negative electron affinity. Therefore, the O-$2p$ subspace exhibits a weak occupation response to local perturbations. Within the LRT framework, this weak response naturally results in a large Hubbard $U_{\mathrm{O}}$~\cite{Gelin:2024, Yu:2014, Moore:2024}. Despite these large $U_{\mathrm{O}}$ values, the inter-site interaction $V_{\mathrm{TM,O}}$ remains small (below 1~eV), indicating that TM--O hybridization has only a minor effect on the electronic structure of these TMOs even when $U_{\mathrm{O}}$ is considered. 
Importantly, the value of $U_{\mathrm{TM}}$ can change when $U_{\mathrm{O}}$ is included in the calculation. Comparing Figs.~\ref{Fig:TMO_wo_O} and \ref{Fig:TMO_with_O}, we observe a slight increase of $U_{\mathrm{TM}}$ for CoO and NiO (by $3$–$5$\%), a substantial increase for FeO (18\%), and almost no change for MnO. Conversely, eACBN0 yields smaller $U_{\mathrm{O}}$ values than LRT ones, in the range of $2-3$~eV. Similar to the case with the $V_{\mathrm{TM,O}}$ correction, the converged Hubbard parameters can vary depending on the Hubbard manifolds being considered. In this self-consistent cycle, the parameters from one iteration feed back into the next as a correction term, thereby influencing the KS states and the final converged values. Comparing Figs.~\ref{Fig:TMO_wo_O} and \ref{Fig:TMO_with_O} for eACBN0, the changes are likewise system dependent and can be small increase, decreases, or negligible variations, depending on whether the DFT+$U$ or DFT+$U$+$V$ scheme is employed.

The effects of $V_{\mathrm{TM,O}}$ and $U_{\mathrm{O}}$ on the magnetic moments are not very significant. The predicted magnetic moments for the TMOs are consistent regardless of computational schemes and they agree well with the experimental range irrespective of inclusion of $V_{\mathrm{TM,O}}$. Conversely, the influence of $V_{\mathrm{TM,O}}$ is pronounced on the band gap. For eACBN0, the inclusion of $V_{\mathrm{TM,O}}$ introduces a non-negligible difference for FeO, CoO, and NiO. Nevertheless, the inclusion of $U_{\mathrm{O}}$ has a greater overall impact on the band gap magnitude than $V_{\mathrm{TM,O}}$ does. 
Without $U_{\mathrm{O}}$ (Fig.~\ref{Fig:TMO_wo_O}), both methods tend to underestimate the band gaps of MnO and NiO, while results for FeO and CoO are generally in good agreement with experiment.  
Including $U_{\mathrm{O}}$ leads to a significant increase in the band gaps within LRT, whereas the corresponding increase within ACBN0 is more moderate (Fig.~\ref{Fig:TMO_with_O}). For MnO, LRT yields a band gap that is close to the experimental values, while eACBN0 still underestimates it. In contrast, for FeO the LRT band gap overshoots the experimental value by approximately $2~\mathrm{eV}$, whereas ACBN0 remains in good agreement with experiment. For CoO, LRT overestimates the lower bound of the experimental band gap range but agrees well with the upper bound, while ACBN0 continues to match the lower bound. Finally, for NiO, LRT overestimates the experimental band gap by about $1-1.5$~eV, whereas ACBN0 falls within the experimental range. 

Finally, following the protocol in Fig.~S1 in the SI~\cite{Timrov:2021}, we compute the self-consistent Hubbard parameters and their corresponding optimized structural parameters using both LRT and eACBN0. The results are presented in Fig.~S2 and~S3 and their discussion is in the SI. For both methods, the structural parameters (lattice constants and rhombohedral angles) are generally in good agreement with available experimental data.  

\subsection*{ZrO$_2$ and BaTiO$_3$}

To further compare the LRT and eACBN0 methods, we examine ZrO$_2$ and BaTiO$_3$, where both Zr$^{4+}$ and Ti$^{4+}$ ions have nominal $d^0$ electronic configurations. Since these materials are nonmagnetic, the accuracy of the predictions will be assessed based on the band gap as well as the structural properties in the case of BaTiO$_3$. 

\begin{figure}[t!]
\centering
\includegraphics[width=1\columnwidth]{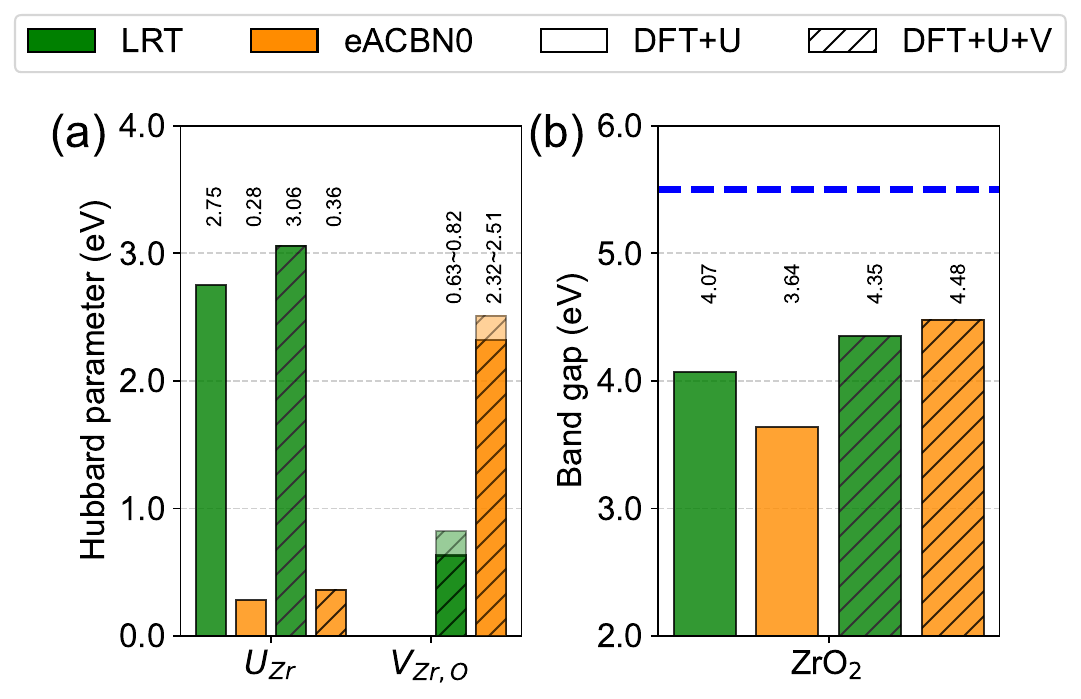}
\caption{DFT+$U$ and DFT+$U$+$V$ self-consistent calculations for ZrO$_{2}$ without on-site corrections on O-$2p$ states ($U_{\mathrm{O}}$). (a) On-site Hubbard $U_{\mathrm{Zr}}$parameter for $4d$ orbitals of Zr ions and inter-site Hubbard $V_{\mathrm{Zr,O}}$ parameter between $4d$ orbitals of Zr ions and $2p$ orbitals of oxygen, and (b) band gap. The experimental band gap~\cite{Bersch:2008} is shown by the dashed blue line in panel (b).}
\label{Fig:ZrO2_wo_O}
\end{figure}

Figure~\ref{Fig:ZrO2_wo_O} shows the results for ZrO$_2$ without the $U_{\mathrm{O}}$ correction. In LRT, the $U$ parameter for Zr-$4d$ states ($U_{\mathrm{Zr}}$) is $\sim 3$~eV, reflecting the finite response arising from the small but nonzero occupancy of Zr-$4d$ states in the PBEsol ground state. In contrast, ACBN0, being more sensitive to the small residual occupancy of the nominal $d^0$ states, yields a much smaller$U_{\mathrm{Zr}}$ value of $\sim 0.3$~eV, which is one order of magnitude smaller than the LRT value. Consequently, the very small $U_{\mathrm{Zr}}$ obtained from ACBN0 leads to a significantly smaller band-gap correction in DFT+$U$ compared to that obtained using LRT [see Fig.~\ref{Fig:ZrO2_wo_O}(b)]. 

For the inter-site interactions between Zr-$4d$ and O-$2p$ states ($V_{\mathrm{Zr,O}}$), LRT yields values below 1 eV, comparable to those obtained for the TMOs discussed in the previous section. Including these interactions in the DFT+$U$+$V$ calculation increases the band gap by 7\% relative to DFT+$U$ with only $U_{\mathrm{Zr}}$ from LRT. In contrast, eACBN0 produces substantially larger $V_{\mathrm{Zr,O}}$ values (2.3–2.5 eV), leading to a 22\% band-gap increase compared to DFT+$U$ with only $U_{\mathrm{Zr}}$, as shown in Fig.~\ref{Fig:ZrO2_wo_O}(b). Since $V_{\mathrm{Zr,O}}$ depends on the interatomic distances between Zr and its nearest-neighbor O atoms, a range of values is reported in Fig.~\ref{Fig:ZrO2_wo_O}(a) rather than a single number. Despite these $V_{\mathrm{Zr,O}}$ corrections, both methods still underestimate the experimental band gap by more than 1 eV~\cite{Bersch:2008}.

\begin{figure}[t!]
\centering
\includegraphics[width=1\columnwidth]{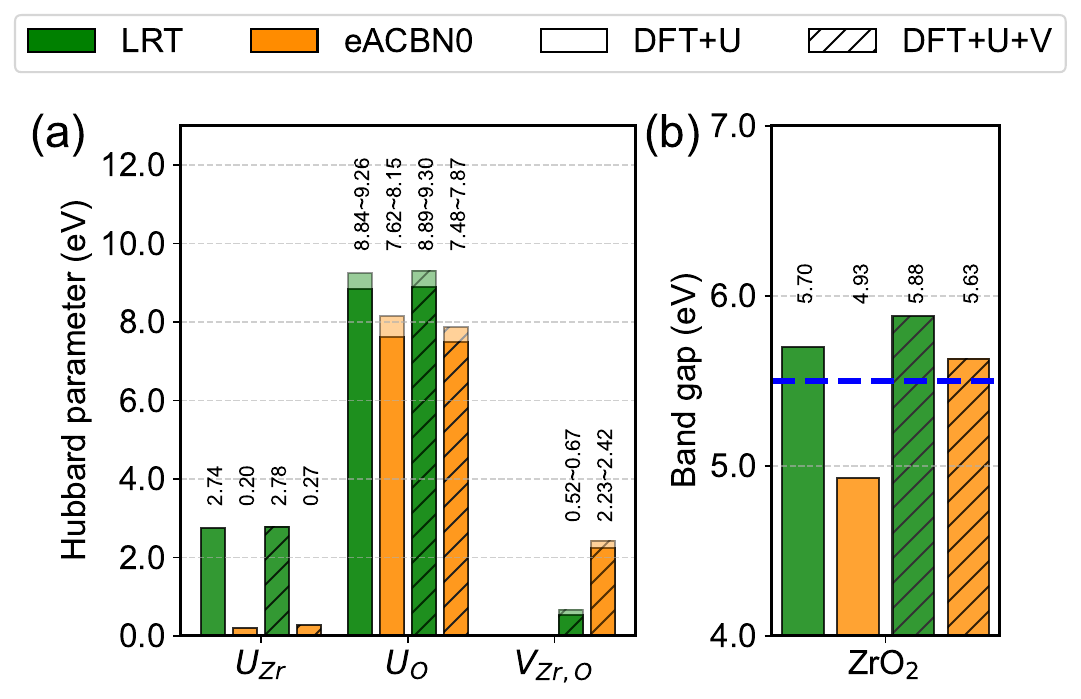}
\caption{DFT+$U$ and DFT+$U$+$V$ self-consistent calculations for ZrO$_{2}$ with on-site $U_{\mathrm{O}}$ corrections on O-$2p$ states. (a) On-site Hubbard $U_{\mathrm{Zr}}$ and $U_{\mathrm{O}}$ parameters for $4d$ orbitals of Zr ions and $2p$ orbitals of oxygen, respectively, and inter-site Hubbard $V_{\mathrm{Zr,O}}$ parameter between $4d$ orbitals of Zr ions and $2p$ orbitals of oxygen, and (b) band gap. The experimental band gap~\cite{Bersch:2008} is shown by the dashed blue line in panel (b).}
\label{Fig:ZrO2_with_O}
\end{figure}

Including $U_{\mathrm{O}}$ substantially increases the band gap, as shown in Fig.~\ref{Fig:ZrO2_with_O}. Both methods (LRT and eACBN0) yield large $U_{\mathrm{O}}$ values in the range of 7–9~eV, with LRT producing the higher values and eACBN0 the lower ones. In the eACBN0 formalism, the two slightly inequivalent O sublattices in ZrO$_2$ are included in the same O-species renormalization factor, while the projections of the KS states onto the local atomic orbitals remain site-dependent and thus retain sensitivity to the distinct local environments. Incorporating these $U_{\mathrm{O}}$ corrections within either DFT+$U$ or DFT+$U$+$V$ consistently increases the band gap relative to calculations without $U_{\mathrm{O}}$, bringing the results closer to the experimental value. Overall, LRT-based predictions (whether or not $V_{\mathrm{Zr,O}}$ is included) tend to overestimate the band gap, whereas eACBN0-based predictions underestimate it within DFT+$U$ and slightly overestimate it within DFT+$U$+$V$.

The fully relaxed structures of ZrO$_2$ are well described by both methods, as shown in Fig.~S4, and they are only weakly affected by the inclusion of $V_{\mathrm{Zr,O}}$ and $U_{\mathrm{O}}$. We also note that the method-dependent variations of the ``structurally self-consistent'' Hubbard parameters during structural relaxation follow the same trends discussed above for the ``self-consistent'' case at a fixed PBEsol geometry.

Next, we turn to BaTiO$_3$, which is similar to ZrO$_2$ in the sense that Ti$^{4+}$ also has a nominal $d^0$ configuration. In this case, we consider not only the band gap but also the crystal-structure properties, which are highly sensitive to Hubbard corrections~\cite{Gebreyesus:2023, Choi:2025}. BaTiO$_3$ is a well-known ferroelectric material, exhibiting a rhombohedral distortion at low temperatures. However, standard DFT+$U$ with sizable $U$ corrections for Ti-$3d$ states incorrectly stabilizes a centrosymmetric cubic structure, thereby suppressing its intrinsic polarization~\cite{Tsunoda2019prm}. However, including the inter-site interactions ($V_{\mathrm{Ti,O}}$) between Ti-$3d$ and O-$2p$ states preserves their hybridization, which is essential for maintaining the rhombohedral distortion and thus restoring the polarization~\cite{Gebreyesus:2023, Choi:2025}.

\begin{figure}[t!]
\centering
\includegraphics[width=1\columnwidth]{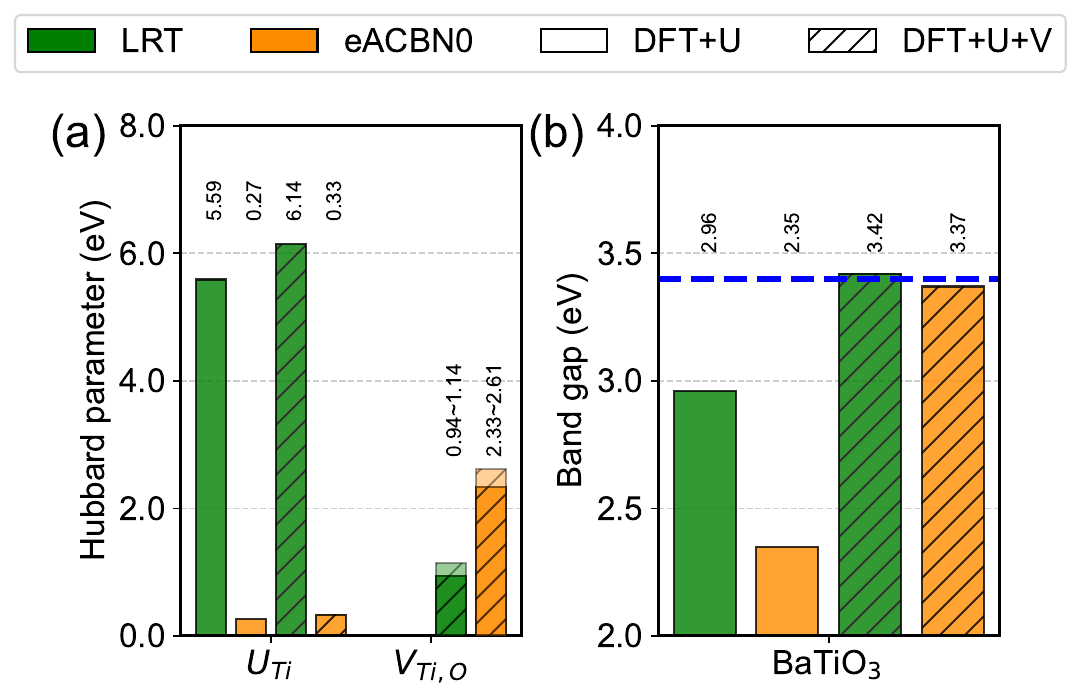}
\caption{DFT+$U$ and DFT+$U$+$V$ self-consistent calculations for BaTiO$_3$ without on-site corrections on O-$2p$ states ($U_{\mathrm{O}}$). (a) On-site Hubbard $U_{\mathrm{Ti}}$ parameter for $3d$ orbitals of Ti ions and inter-site Hubbard $V_{\mathrm{Ti,O}}$ parameter between $3d$ orbitals of Ti ions and $2p$ orbitals of oxygen, and (b) band gap. The experimental band gap~\cite{Wemple:1970} is shown by the dashed blue line in panel (b).}
\label{Fig:BaTiO3_wo_O}
\end{figure}

In Fig.~\ref{Fig:BaTiO3_wo_O} we present the results obtained using the ``self-consistent'' scheme described in Fig.~\ref{Fig:flowchart} without the $U_{\mathrm{O}}$ correction. Within LRT, the Hubbard parameter for the Ti-$3d$ states ($U_{\mathrm{Ti}}$) is $\sim 6$~eV, reflecting the finite response associated with the small but nonzero $3d$ occupancy in the PBEsol ground state, similarly to the behavior observed in ZrO$_2$. Since Ti-$3d$ states are more localized than Zr-$4d$ states, the resulting $U_{\mathrm{Ti}}$ is roughly twice as large as $U_{\mathrm{Zr}}$ in LRT. In contrast, ACBN0 yields a very small $U_{\mathrm{Ti}}$ of $\sim 0.3$~eV, identical to $U_{\mathrm{Zr}}$, since the Ti-$3d$ and Zr-$4d$ occupancies are both very low. Thus, ACBN0 appears largely insensitive to the degree of $d$-state localization ($3d$ vs.\ $4d$), depending primarily on the actual orbital occupancies. As shown in Fig.~\ref{Fig:BaTiO3_wo_O}, these differences lead to markedly different DFT+$U$ outcomes: LRT-based DFT+$U$ underestimates the band gap by $\sim 13$\%, whereas eACBN0-based DFT+$U$ underestimates it by $\sim 31$\%. This picture changes substantially when $V_{\mathrm{Ti,O}}$ is included. The $V_{\mathrm{Ti,O}}$ values obtained from LRT and eACBN0 are similar in magnitude to those found for ZrO$_2$. With these inter-site interactions included, both methods yield band gaps in excellent agreement with experiment within DFT+$U$+$V$. Thus, an accurate band gap is achieved either through a combination of a large $U_{\mathrm{Ti}}$ and small $V_{\mathrm{Ti,O}}$ (as in LRT) or a small $U_{\mathrm{Ti}}$ and large $V_{\mathrm{Ti,O}}$ (as in eACBN0).

\begin{figure}[t!]
\centering
\includegraphics[width=1\columnwidth]{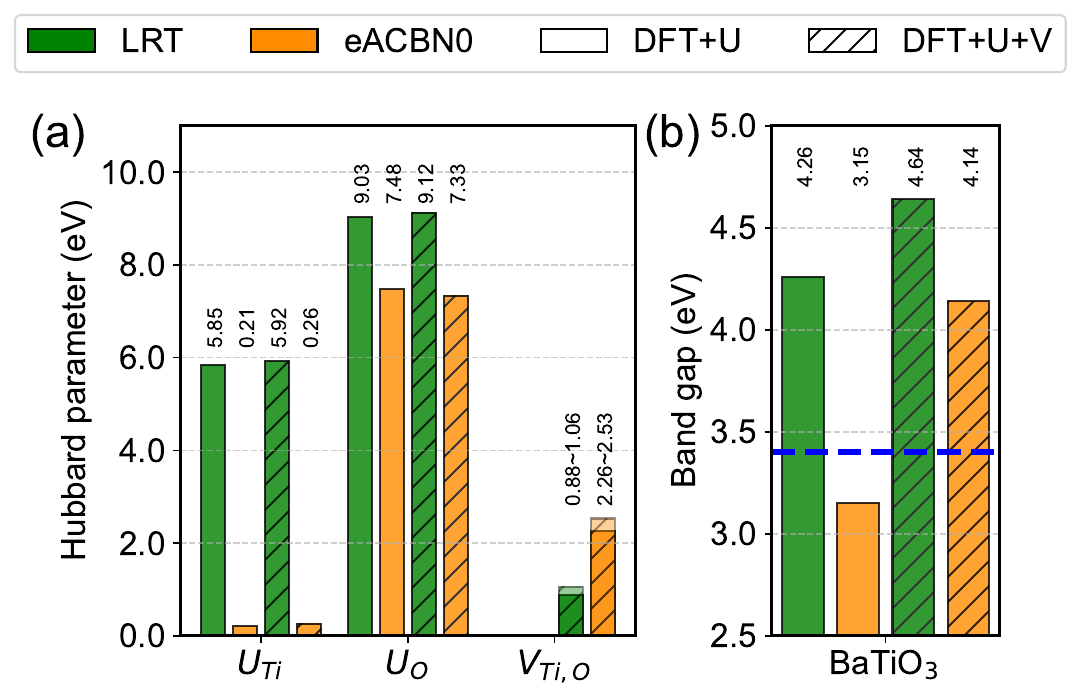}
\caption{DFT+$U$ and DFT+$U$+$V$ self-consistent calculations for BaTiO$_3$ with on-site $U_{\mathrm{O}}$ corrections on O-$2p$ states. (a) On-site Hubbard $U_{\mathrm{Ti}}$ and $U_{\mathrm{O}}$ parameters for $3d$ orbitals of Ti ions and $2p$ orbitals of oxygen, respectively, and inter-site Hubbard $V_{\mathrm{Ti,O}}$ parameter between $3d$ orbitals of Ti ions and $2p$ orbitals of oxygen, and (b) band gap. The experimental band gap~\cite{Wemple:1970} is shown by the dashed blue line in panel (b).}
\label{Fig:BaTiO3_with_O}
\end{figure}

As shown in Fig.~\ref{Fig:BaTiO3_with_O}(a), the $U_{\mathrm{O}}$ values obtained from LRT and eACBN0 are comparable to those for ZrO$_2$ (see Fig.~\ref{Fig:ZrO2_with_O}(a)). Including these $U_{\mathrm{O}}$ corrections has a strong impact on the band gap of BaTiO$_3$. In particular, when $U$ corrections are applied to both Ti-$3d$ and O-$2p$ states (and even when the inter-site term $V_{\mathrm{Ti,O}}$ is also included) both methods substantially overestimate the band gap (Fig.~\ref{Fig:BaTiO3_with_O}(b)). The only exception is the DFT+$U$ calculation with ACBN0-derived $U$ values, which still underestimates the band gap but by a smaller margin (7\%).

Figure~S5 in the SI shows the results of the ``structurally self-consistent'' calculations for BaTiO$_3$. The Hubbard parameters $U_{\mathrm{Ti}}$ and $V_{\mathrm{Ti,O}}$ are only slightly modified compared with those obtained using the PBEsol geometry (compare Fig.~S5(c) with Fig.~\ref{Fig:BaTiO3_wo_O}(a)). Without $U_{\mathrm{O}}$, the two methods exhibit different structural trends: LRT-based DFT+$U$ predicts a cubic phase ($\alpha = 90^\circ$), whereas eACBN0-based DFT+$U$ retains the rhombohedral distortion ($\alpha < 90^\circ$), as shown in Fig.~S5(b). When $V_{\mathrm{Ti,O}}$ is included, both methods capture the rhombohedral distortion, but their quantitative predictions differ. Specifically, DFT+$U$+$V$ based on LRT parameters yields a lattice parameter, rhombohedral angle, and band gap in very good agreement with experiment, whereas DFT+$U$+$V$ based on eACBN0 parameters shows significantly larger deviations for all three properties.

When $U_{\mathrm{O}}$ is included, the structural behavior changes, as shown in Figs.~S5(e)–(h). In this case, LRT fails to capture the ferroelectric nature of BaTiO$_3$: the large $U_{\mathrm{Ti}}$ and $U_{\mathrm{O}}$ values overly localizes the Ti-$3d$ and O-$2p$ states, respectively, driving the system into a non-polar cubic phase even when the small inter-site interaction $V_{\mathrm{Ti,O}}$ is added (Fig.~S5(f)). A similar trend occurs in eACBN0-based DFT+$U$ calculations, where the combined inclusion of small $U_{\mathrm{Ti}}$ and large $U_{\mathrm{O}}$ suppresses the polarization and restores cubic symmetry. However, unlike LRT, adding large $V_{\mathrm{Ti,O}}$ within the eACBN0 framework recovers the rhombohedral distortion and the associated polarization, consistent with previous findings~\cite{Choi:2025}. Nonetheless, the resulting lattice parameter and band gap remain substantially overestimated.

These results highlight that, although both approaches rely on similar Hubbard correction schemes, they lead to qualitatively different effects on structural and electronic properties of BaTiO$_3$. Namely, LRT yields the most accurate properties when $U_{\mathrm{Ti}}$ and $V_{\mathrm{Ti,O}}$ are included but $U_{\mathrm{O}}$ is excluded, whereas eACBN0 achieves its best accuracy when all correction terms are included.

\subsection*{ZnO}

After examining TMOs with partially filled $d$ states (from Mn$^{2+}$ to Ni$^{2+}$) and then early TM compounds with nominal $d^0$ configurations (Ti$^{4+}$ and Zr$^{4+}$), we now turn to a material with a late $d$-block element: ZnO, where Zn$^{2+}$ has a nominally filled $d^{10}$ configuration. ZnO represents the opposite extreme of the $d$-electron occupancy spectrum relative to the $d^0$ systems, offering a distinctly different test case for evaluating Hubbard parameters.

\begin{figure}[t!]
\centering
\includegraphics[width=1\columnwidth]{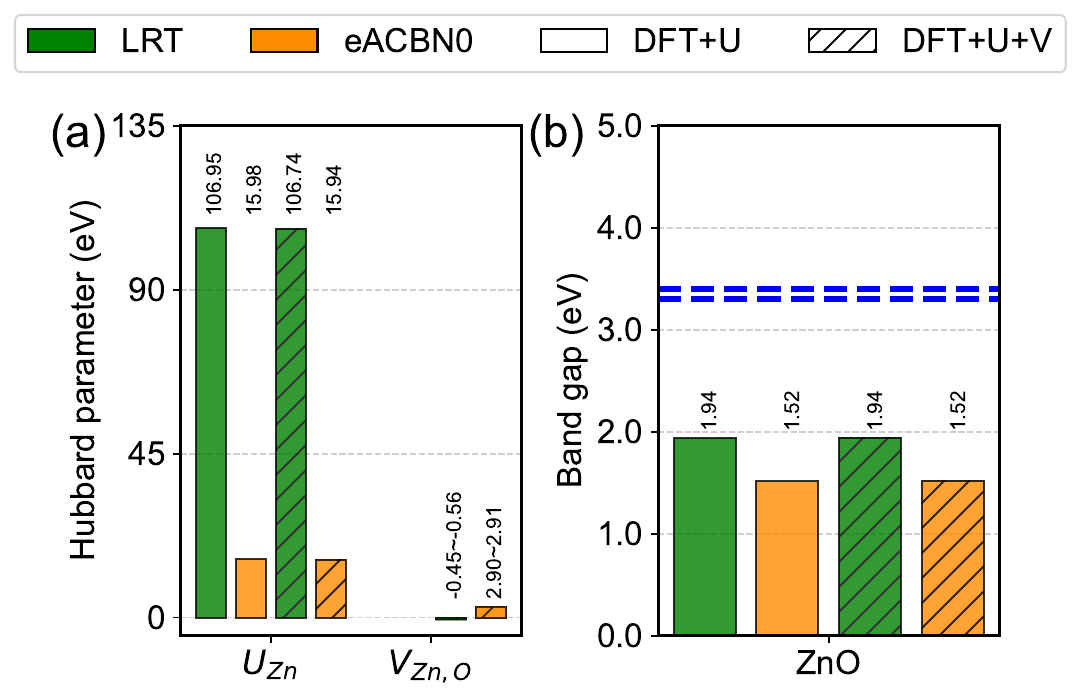}
\caption{DFT+$U$ and DFT+$U$+$V$ self-consistent calculations for ZnO without on-site corrections on O-$2p$ states ($U_{\mathrm{O}}$). (a) On-site Hubbard $U_{\mathrm{Zn}}$ parameter for $3d$ orbitals of Zn ions and inter-site Hubbard $V_{\mathrm{Zn,O}}$ parameter between $3d$ orbitals of Zn ions and $2p$ orbitals of oxygen, and (b) band gap. The experimental band gap~\cite{Dong2004prb} is shown by the dashed blue lines in panel (b).}
\label{Fig:ZnO_wo_O}
\end{figure}

We first discuss the results for ZnO without the $U_{\mathrm{O}}$ correction. As shown in Fig.~\ref{Fig:ZnO_wo_O}, the LRT method yields an unphysically large self-consistent $U_{\mathrm{Zn}}$ exceeding 100~eV. Even without performing the full self-consistent cycle (see Fig.~\ref{Fig:flowchart}), i.e. using the ``one-shot'' scheme, the LRT estimate already exceeds 25~eV. This well-known limitation of LRT arises from the fully occupied Zn-$3d$ shell, which leads to spuriously large $U$ values~\cite{Yu:2014}. The “structurally self-consistent’’ scheme produces a similarly large $U_{\mathrm{Zn}}$, as shown in Fig.~S6. It is important to emphasize that in ZnO the valence-band maximum is primarily O-$2p$ in character, while the conduction-band minimum is dominated by Zn-$4s$ states. As a result, applying $U_{\mathrm{Zn}}$ to the Zn-$3d$ states influences the band gap only indirectly. In fact, several studies have emphasized that applying a Hubbard correction to the Zn-$4s$ subspace can be more effective than correcting only the deep Zn-$3d$ states~\cite{Paudel2008prb,Lambert2023prr}. Related evidence that SIEs in metal $4s$ orbitals can be non-negligible has also been reported in molecular systems~\cite{Kulik2010jcp}. Indeed, LRT-based DFT+$U$ increases the PBEsol band gap from $\sim 0.7$~eV to 1.94~eV, which remains well below the experimental value of $3.3$~eV~\cite{Dong2004prb}. In contrast, the eACBN0 approach predicts also large but more moderate $U_{\mathrm{Zn}}$ of $\sim 16$~eV, resulting in a slightly smaller band gap of 1.52~eV, as shown in Fig.~\ref{Fig:ZnO_wo_O}(b).

For the inter-site interactions between Zn-$3d$ and O-$2p$ states ($V_{\mathrm{Zn,O}}$), LRT yields small negative values. These negative values likely originate from numerical instabilities during the inversion of response matrices whose elements are close to zero and therefore highly sensitive to numerical noise. In contrast, eACBN0 produces significantly larger and more stable $V_{\mathrm{Zn,O}}$ values of about 3~eV, consistently with the trends observed for the other materials discussed above. However, these inter-site corrections have no effect on the band gap, as shown in Fig.~\ref{Fig:ZnO_wo_O}(b). This insensitivity arises because the band gap in ZnO is defined between O-$2p$ (valence) and Zn-$4s$ (conduction) states, while the Zn-$3d$ states lie much deeper in energy. As a result, the $V_{\mathrm{Zn,O}}$ term mainly alters the hybridization between the O-$2p$ and Zn-$3d$ states but does not influence the band edges, leaving the band gap unchanged.

\begin{figure}[t!]
\centering
\includegraphics[width=1\columnwidth]{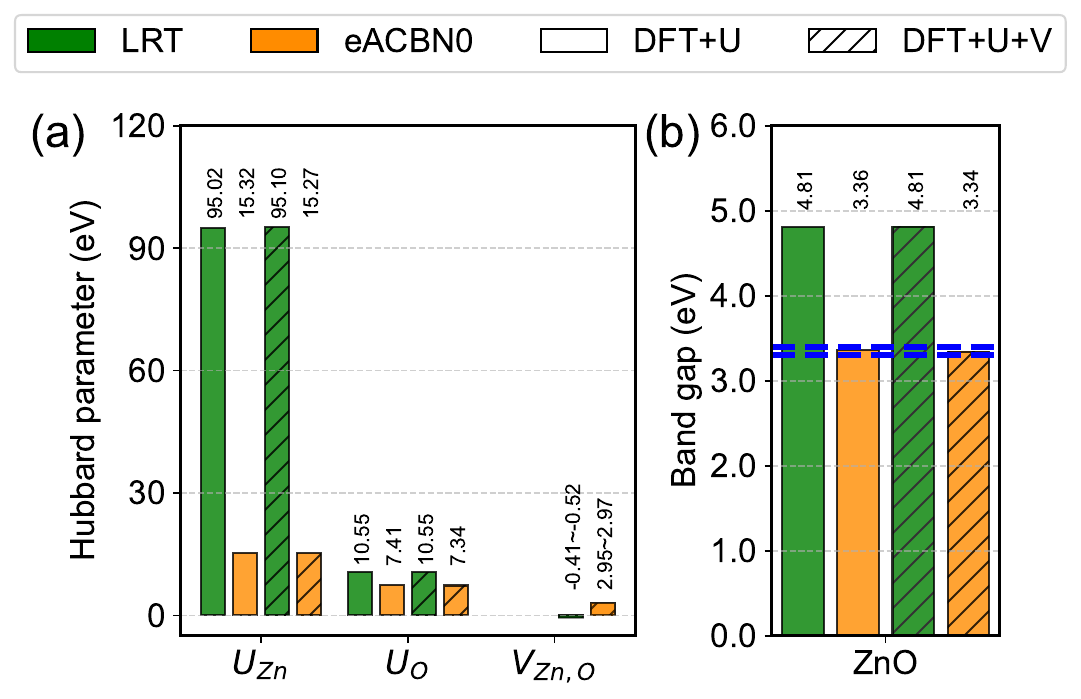}
\caption{DFT+$U$ and DFT+$U$+$V$ self-consistent calculations for ZnO with on-site $U_{\mathrm{O}}$ corrections on O-$2p$ states. (a) On-site Hubbard $U_{\mathrm{Zn}}$ and $U_{\mathrm{O}}$ parameters for $3d$ orbitals of Zn ions and $2p$ orbitals of oxygen, respectively, and inter-site Hubbard $V_{\mathrm{Zn,O}}$ parameter between $3d$ orbitals of Zn ions and $2p$ orbitals of oxygen, and (b) band gap. The experimental band gap~\cite{Dong2004prb} is shown by the dashed blue lines in panel (b).}
\label{Fig:ZnO_with_O}
\end{figure}

Including $U_{\mathrm{O}}$ in addition to $U_{\mathrm{Zn}}$, as shown in Fig.~\ref{Fig:ZnO_with_O}, leads to markedly different outcomes. In LRT, the already unphysical $U_{\mathrm{Zn}}$ value combined with a large $U_\mathrm{O} \sim 11$~eV results in a band gap that overestimates the experimental value by 44\%. In contrast, eACBN0 yields still large but more moderate values, $U_{\mathrm{Zn}} \sim 15$~eV and $U_\mathrm{O} \sim 7$~eV, which together produce a band gap in excellent agreement with experiment. As before, adding the inter-site term $V_{\mathrm{Zn,O}}$ has no influence on the band gap for the reasons discussed above. These results show that the band gap is highly sensitive to $U_{\mathrm{O}}$, since this correction directly shifts the O-$2p$ states that form the valence-band maximum in ZnO.

When structural relaxations are included, both the Hubbard parameters and the resulting band gaps follow similar trends as described above (see Fig.~S6). We also note that, with these Hubbard corrections, both methods reproduce the experimental structural parameters very well (see Fig.~S6(a) and (d)).

\section*{DISCUSSION}

Based on the numerical results presented in the previous section, we now provide a discussion and interpretation of the observed trends in the behavior of the on-site $U$ and inter-site $V$ Hubbard parameters obtained from LRT and eACBN0. Despite the fundamentally different mathematical formulations and physical concepts underlying these two methods, there are cases where they yield comparable $U$ values, as well as cases where the results diverge dramatically. By contrast, the inter-site $V$ parameters systematically exhibit large discrepancies between the two approaches. In what follows, we analyze these similarities and differences in more detail.

We start with the on-site $U_{\mathrm{TM}}$ values. For TMOs with partially occupied $3d$ shells, LRT and eACBN0 predict similar $U_{\mathrm{TM}}$ values except $U$ for Fe in FeO that differs by $2$–$3$ eV. 
This discrepancy can be largely attributed to the choice of pseudopotentials, which determine the Hubbard projectors and can significantly influence the computed $U_{\mathrm{TM}}$ values, particularly in LRT. As a result, the level of agreement between LRT and eACBN0 can vary depending on the degree of ionization and on the specific construction scheme of the pseudopotentials.
However, when the $3d$ shell is nearly empty or fully occupied, the two methods behave in fundamentally different ways. In particular, for nominally empty $3d$ states, LRT still produces finite and sizable $U_{\mathrm{TM}}$ values in the range of $3$–$6$ eV, while for fully occupied $3d$ states it yields unphysically large values. The latter is a well-known limitation of the LRT framework~\cite{Yu:2014}: for filled shells, the occupation matrices exhibit vanishing variations under a local electronic perturbation, leading to nearly singular response matrices whose inversion generates spuriously large $U_{\mathrm{TM}}$ values. In ZnO, this behavior can also be interpreted as an indication that the Zn-$3d$ states are not the most relevant one for the band edge, since the conduction band minimum has predominantly Zn-$4s$ character. Consequently, applying Hubbard corrections directly to the Zn-$4s$ manifold has been suggested as a useful strategy to better target the states controlling the band edges in ZnO~\cite{Paudel2008prb,Lambert2023prr}. The behavior of eACBN0 is markedly different: It produces $U_{\mathrm{TM}}$ values that are finite and in the range of physical values even for closed shells. The root cause of this behavior lies in the definition of the auxiliary objects - the renormalized occupation numbers [Eq.~\eqref{eq:rom}] and renormalized density matrices [Eq.~\eqref{eq:rdm}]. These quantities are constructed to ensure that the sum of the Mulliken charge of each atom and the interatomic charge reproduces the correct total electronic population. As a consequence, the Hubbard parameters derived from eACBN0 naturally become very small for nearly empty shells, while for fully filled $3d$ shells they approach values of $\sim 16$~eV. This trend is not mirrored in LRT because of the different physical idea: $U_{\mathrm{TM}}$ is computed from the occupation-matrix variations in response to local electronic perturbations. For nearly empty shells, even an infinitesimal local perturbation can induce relatively large fractional changes in the occupation matrix, which translates into sizable $U_{\mathrm{TM}}$ values. Conversely, for fully occupied shells, the occupation matrix is extremely rigid, and the small changes captured by the linear-response equations produce nearly vanishing matrix elements. The inversion of these nearly singular matrices amplifies the response artificially, resulting in unphysically large $U_{\mathrm{TM}}$ values. Thus, while LRT and eACBN0 can yield comparable results for partially filled shells, their behavior diverges sharply at the two extremes of the $d$-electron occupation spectrum.

Regarding the on-site Hubbard $U_{\mathrm{O}}$ parameter for O-$2p$ states, as obtained from LRT and eACBN0, we observe notable differences. On the one hand, LRT consistently predicts large $U_{\mathrm{O}}$ values, close to $10$~eV, essentially independent of the specific compound under investigation. This is consistent with the intrinsically large chemical hardness of oxygen, which makes the O-$2p$ subspace weakly respond to the local perturbations and relatively insensitive to the chemical environment, as discussed by Moore \textit{et al.}~\cite{Moore:2024} In contrast, the eACBN0 approach provides a much broader and more system-dependent distribution of $U_{\mathrm{O}}$ values, ranging from $2$ to $8$~eV. Specifically, $U_{\mathrm{O}}$ is about $2$–$3$~eV for TMOs, but increases to $7 - 8$~eV in ZnO, ZrO$_2$ and BaTiO$_3$. These results indicate that the $U_{\mathrm{O}}$ value does not follow the same trend as the $U_{\mathrm{TM}}$ values, which generally scale with the electron count within eACBN0. This behavior can be rationalized by considering that the degree of localization and hybridization of O-$2p$ states varies strongly across materials. In some cases, such as in ZrO$_2$ and BaTiO$_3$, the oxygen states are more localized, resulting in larger $U_{\mathrm{O}}$ values. Additionally, it is important to account for the fact that O-$2p$ states participate in significant charge redistribution and hybridization with TM-$3d$. This naturally calls for the inclusion of inter-site $V$ terms in order to properly capture the covalency of such interactions, since purely on-site corrections are insufficient to accurately describe the electronic interactions.

Next, we turn to the inter-site $V$ interactions between TM-$d$ and O-$p$ states in the studied materials. From our LRT calculations, we find that for all compounds except ZnO, $V$ values are typically around $1$~eV or smaller, while in ZnO they are small and negative. Such values are characteristic of LRT: $V$ corresponds to the off-diagonal elements of the response matrix, describing how the occupation on a given atom changes when a neighboring atom is perturbed. Since these inter-site responses are much weaker than the on-site ones (which define $U$), the resulting $V$ is generally small. In ZnO, the small negative $V$ originates from numerical instabilities caused by inverting nearly singular response matrices. In contrast, eACBN0 yields a markedly different picture: The inter-site $V$ values are much larger, often comparable in magnitude to the on-site $U$. For instance, in TMOs with partially occupied $3d$ shells and in ZnO, $V$ is $2$–$5$ times smaller than the corresponding $U_{\mathrm{TM}}$. Even more strikingly, in ZrO$_2$ and BaTiO$_3$, $V$ exceeds the $U_{\mathrm{TM}}$ values of the nominally empty $d$ states by a factor of $\sim 10$. This behavior stems from the fact that in eACBN0, the formation of a chemical bond involves charge redistribution between atoms, placing $U$ and $V$ on a more equal footing than in LRT. Indeed, Eq.~\eqref{eq:rom} shows that $V^{IJ}$ scales with the sum of Mulliken charges of atoms $I$ and $J$, and is therefore not intrinsically smaller than its on-site counterpart. Moreover, since eACBN0 derives $U$ and $V$ from energy contributions that depend on the spatial distribution of electronic charges, bonding tends to lower on-site $U$ while enhancing inter-site $V$. In this framework, the ratio $U/V$ reflects the balance between atomic and bond-centered charges, and is thus directly tied to bond strength. Consequently, $V$ remains sizable relative to $U$ in all systems. By contrast, in LRT the $U$ and $V$ parameters emerge from occupation matrix responses, where the dominant contribution is always local. As a result, $U$ is much larger than $V$, with ratios often an order of magnitude.

Finally, we discuss the sensitivity of the band gaps to the Hubbard parameters obtained from LRT and eACBN0. For FeO and CoO, applying $U$ on the TM ions alone yields band gaps in good agreement with experiment, whereas for MnO and NiO the gaps are somewhat underestimated. In these cases, the inter-site $V$ contribution is negligible. The inclusion of $U$ on oxygen strongly affects the predicted gaps. LRT tends to overestimate the experimental values for FeO and NiO, while yielding good agreement for MnO and for the upper bound of the experimental band gap range in CoO. Conversely, eACBN0 shows good agreement with experimental values for FeO and NiO, as well as with the lower bound of the CoO band gap, while significantly underestimating the band gap of MnO. These trends directly reflect the substantial differences in the values of $U$ on oxygen obtained from the two methods. For ZrO$_2$, applying $U$ only on the TM ion underestimates the band gap for both LRT and eACBN0, and the inclusion of $V$ has only a small effect. Adding $U$ on oxygen generally increases the gap, but this often leads to an overestimation relative to experiment. For BaTiO$_3$, applying $U$ only on the TM ion while including $V$ produces accurate band gaps for both methods, whereas adding $U$ on oxygen significantly overestimates the gaps. For ZnO, applying $U$ on TM ions (with or without $V$) underestimates the band gaps in both LRT and eACBN0. Including $U$ on oxygen, however, overshoots the gaps in LRT while eACBN0 yields results in good agreement with experiment. Overall, these trends indicate that the predicted band gaps are most sensitive to the treatment of oxygen states, while inter-site $V$ typically has a smaller effect. In particular, LRT tends to produce larger $U_{\mathrm{O}}$ values than eACBN0, often leading to overestimation of the gaps when oxygen is corrected. Conversely, eACBN0 often provides a more balanced description across different materials. These observations suggest that careful consideration of which Hubbard terms to include (especially on oxygen) can provide practical guidance for achieving accurate band gaps in TM compounds, highlighting a pathway for improved computational predictions in related materials.

In summary, we have carried out a systematic comparison of two widely used approaches, LRT and eACBN0, for computing on-site $U$ and inter-site $V$ Hubbard parameters across three classes of materials spanning nearly empty, partially filled, and fully occupied TM-$d$ shells. These two methods rest on fundamentally different physical pictures and mathematical frameworks, yet our analysis reveals both points of agreement and distinct discrepancies. For TMOs with partially occupied $d$ states, the two methods yield broadly consistent $U$ values, whereas for nearly empty or fully filled $d$ shells the disagreement becomes substantial, reflecting intrinsic differences in how each method treats electronic interactions. For the O-$2p$ states, LRT systematically predicts large $U$ values independent of chemistry, while eACBN0 produces a much wider spread, sensitive to localization and hybridization with TM-$d$ states. The contrast is even sharper for the inter-site $V$: LRT consistently yields small values ($\sim$1 eV or less), while eACBN0 predicts much larger values, often comparable to on-site $U$. These differences can be traced to the distinct mathematical assumptions underpinning the definitions of $U$ and $V$, and to the different physical insights each method encodes - LRT focusing on response functions, eACBN0 on energetics and charge redistribution. 
Since the Hubbard parameters obtained from these approaches are ultimately used within the same DFT+$U$ or DFT+$U$+$V$ energy functional, they are often implicitly assumed to represent the same physical quantity. However, the differences observed here suggest that their interpretation may, at least in part, remain method dependent, particularly in how the screening processes are captured for a given electronic configuration.
Overall, our study highlights that while certain parallels exist for the considered set of materials, the divergences between LRT and eACBN0 are pronounced and can affect the predicted material properties for specific cases. This work represents a first step toward a systematic understanding of how these methods compare and where they diverge. Bridging their conceptual and mathematical foundations, and testing them across a broader range of materials and chemistry, remains an important direction for future developments.

\section*{METHODS}
\subsection*{Computational details}
All calculations were performed using the plane-wave pseudopotential method as implemented in the \textsc{Quantum ESPRESSO} (QE) distribution~\cite{Giannozzi:2009, Giannozzi:2017, Giannozzi:2020}. For the xc functional, we adopted the $\sigma$-GGA with the PBEsol parametrization~\cite{Perdew:2008}, and used pseudopotentials from the GBRV library (v1.5)~\cite{Garrity:2014}. The KS wavefunctions and potentials were expanded in plane waves with kinetic-energy cutoffs of 65 and 780~Ry for the TMOs, and 60 and 600~Ry for ZnO, ZrO$_2$, and BaTiO$_3$. The first BZ was sampled using uniform $\Gamma$-centered $\mathbf{k}$-point meshes: 10$\times$10$\times$10 for TMOs, 12$\times$12$\times$7 for ZnO, 7$\times$7$\times$6 for ZrO$_2$, and 8$\times$8$\times$8 for BaTiO$_3$. Spin-orbit coupling was neglected. Geometry optimizations were carried out using the Broyden–Fletcher–Goldfarb–Shanno (BFGS) algorithm~\cite{Fletcher:1987}, with convergence criteria of $10^{-4}$~Ry for the total energy, $10^{-3}$~Ry/Bohr for forces, and $0.5$~kbar for pressure in TMOs. For ZnO, ZrO$_2$, and BaTiO$_3$, criteria of $10^{-6}$~Ry for the total energy, $10^{-5}$~Ry/Bohr for forces, and $0.5$~kbar for pressure were adopted. For antiferromagnetic TMOs, we estimated the magnetic moment per TM ion from the absolute magnetization at the end of the self-consistent calculations. Since the spin polarization is largely localized on the TM sites, we computed the moment per ion by dividing the converged absolute magnetization of the unit cell by the number of TM atoms.

The LRT Hubbard parameters were computed using DFPT~\cite{Timrov:2018, Timrov:2021} as implemented in the \textsc{HP} code~\cite{Timrov:2022}. Hubbard projectors were constructed from atomic orbitals orthogonalized using the L\"owdin scheme~\cite{Lowdin:1950}. In this scheme, the L\"owdin orthogonalization is applied to the full set of atomic orbitals provided by the pseudopotentials, rather than being restricted to the selected Hubbard manifolds. 

Uniform $\Gamma$-centered $\mathbf{k}$- and $\mathbf{q}$-point grids were employed, with sizes of 10$\times$10$\times$10 and 6$\times$6$\times$6 for TMOs, 12$\times$12$\times$7 and 6$\times$6$\times$3 for ZnO, 7$\times$7$\times$6 and 3$\times$3$\times$3 for ZrO$_2$, and 8$\times$8$\times$8 and 4$\times$4$\times$4 for BaTiO$_3$, respectively. 

The eACBN0 Hubbard parameters were computed using our modified in-house QE package~\cite{Lee:2020,Yang:2021}. For consistency with the LRT approach, the Hubbard projectors were constructed from atomic orbitals orthogonalized using the same L\"owdin scheme~\cite{Lowdin:1950}. The same $\Gamma$-centered $\mathbf{k}$-point meshes used for the electronic structure calculations were employed, and convergence thresholds of $10^{-10}$ and $10^{-15}$~Ry were applied for the total eACBN0 energy and SCF cycle, respectively.

The $U$ and $V$ parameters were obtained using two self-consistent protocols, summarized in Fig.~\ref{Fig:flowchart} and Fig.~S1 of the SI. In the main text, we report the values computed by iteratively evaluating $U$ and $V$ on the PBEsol-relaxed structures using the LRT and eACBN0 methods until convergence (the ``self-consistent'' scheme). In the case of ZnO, however, this iterative procedure was limited to two steps due to its tendency to diverge~\cite{Yu:2014}. For completeness, the SI also presents results from the ``structurally self-consistent'' scheme, in which DFT+$U$+$V$ structural optimizations (including Hubbard contributions to forces and stresses~\cite{Timrov:2020b}) are performed, and the Hubbard parameters are recomputed until both the structure and the $U$ and $V$ values are converged. Moreover, for FeO and CoO, we employ occupation matrix control~\cite{Allen:2014} by initializing the $3d$ occupation matrices with $3d^6$ and $3d^7$ configurations for Fe and Co, respectively. These occupations are appropriately distributed between spin-up and spin-down channels with integer eigenvalues. This procedure is necessary to ensure convergence to the correct insulating ground state~\cite{Cococcioni:2005}. Without it, the systems tend to converge to a metallic solution, leading to different values of the Hubbard parameters.


\section*{DATA AVAILABILITY}

The data used to produce the results of this work is available in the Materials Cloud Archive~\cite{MaterialsCloudArchive2026}. Our modified in-house version of \textsc{Quantum ESPRESSO}, used for eACBN0 calculations, is available at \url{https://github.com/KIAS-CMT/DFT-U-V}. All LRT calculations, in contrast, were performed using the official \textsc{Quantum ESPRESSO} distribution, which can be accessed at \url{https://www.quantum-espresso.org/}


\section*{ACKNOWLEDGEMENTS}

We thank Alberto Carta for fruitful discussions.
W.Y. was supported by KIAS individual Grant (No. QP090102).
I.T. acknowledges support from the Swiss National Science Foundation (Grant No.~200021-227641 and No.~200021-236507).
Y.-W.S. was supported by KIAS individual Grant (No. CG031509).
Computer time was provided by the Swiss National Supercomputing Centre (CSCS) under project No.~s1073, lp18, s1326, s1335 and the CAC of KIAS.


\bibliography{biblio}

@article{Allen:2014,
	title = {{Occupation matrix control of d- and f-electron localisations using {DFT} + {U}}},
	volume = {16},
	doi = {10.1039/C4CP01083C},
	journal = {Phys. Chem. Chem. Phys.},
	author = {Allen, Jeremy P. and Watson, Graeme W.},
	year = {2014},
	pages = {21016},
}

@article{Yang:2025,
  title = {First-principles electron-phonon interactions with self-consistent Hubbard interaction: Application to transparent conducting oxides},
  author = {Yang, Wooil and Tiwari, Sabyasachi and Giustino, Feliciano and Son, Young-Woo},
  journal = {Phys. Rev. B},
  volume = {112},
  issue = {7},
  pages = {075203},
  numpages = {15},
  year = {2025},
  month = {Aug},
  publisher = {American Physical Society},
  doi = {10.1103/w2y5-rl8s}
}

@article{anisimov:1991,
  author  = {V.I.~Anisimov and J.~Zaanen and O.K.~Andersen},
  title   = {{Band theory and Mott insulators: Hubbard $U$ instead of Stoner $I$}},
  journal = {Phys. Rev. B},
  volume  = {44},
  pages   = {943},
  year    = {1991},
  DOI = {10.1103/PhysRevB.44.943}
}

@article{Aykol:2014,
  author  = {Aykol, M. and Wolverton, C.},
  title   = {{Local environment dependent GGA+$U$ method for accurate thermochemistry of transition metal compounds}},
  journal = {Phys. Rev. B},
  volume  = {90},
  pages   = {115105},
  year    = {2014},
  DOI = {10.1103/PhysRevB.90.115105}
}

@article{Adamo:1999,
  author  = {Adamo, C. and Barone, V.},
  title   = {{Toward reliable density functional methods without adjustable parameters: The PBE0 model}},
  journal = {J. Chem. Phys.},
  volume  = {110},
  pages   = {6158},
  year    = {1999},
  DOI = {10.1063/1.478522}
}

@article{Artrith:2022,
  author  = {Artrith, N. and Torres, J.A.G. and Urban, A. and Hybertsen, M.S.},
  title   = {{Data-driven approach to parameterize SCAN+$U$ for an accurate description of $3d$ transition metal oxide thermochemistry}},
  journal = {Phys. Rev. Materials},
  volume  = {6},
  pages   = {035003},
  year    = {2022},
  DOI = {10.1103/PhysRevMaterials.6.035003}
}

@article{Zhou:2004b,
  author  = {Zhou, F. and Cococcioni, M. and Marianetti, C.A. and Morgan, D. and Ceder, G.},
  title   = {{First-principles prediction of redox potentials in transition-metal compounds with LDA+$U$}},
  journal = {Phys. Rev. B},
  volume  = {70},
  pages   = {235121},
  year    = {2004},
  DOI = {10.1103/PhysRevB.70.235121}
}

@article{Heyd:2003,
  author  = {J.~Heyd and G.E.~Scuseria and M.~Ernzerhof},
  title   = {{Hybrid functionals based on a screened Coulomb potential}},
  journal = {J. Chem. Phys.},
  volume  = {118},
  pages   = {8207},
  year    = {2003},
  DOI = {10.1063/1.1564060}
}

@article{Hedin:1965,
  author  = {Hedin, L.},
  title   = {{New Method for Calculating the One-Particle Green's Function with Application to the Electron-Gas Problem}},
  journal = {Phys. Rev.},
  volume  = {139},
  pages   = {A796},
  year    = {1965},
  doi = {10.1103/PhysRev.139.A796},
}

@article{Himmetoglu:2011,
  author  = {B.~Himmetoglu and R.M.~Wentzcovitch and M.~Cococcioni},
  title = {{First-principles study of electronic and structural properties of CuO}},
  journal = {Phys. Rev. B},
  volume  = {84},
  pages   = {115108},
  year    = {2011},
  doi = {10.1103/PhysRevB.84.115108},
}

@article{Himmetoglu:2014,
  author  = {B.~Himmetoglu and A.~Floris and S. de Gironcoli and M.~Cococcioni},
  title   = {{Hubbard-corrected DFT energy functionals: The LDA+U description of correlated systems}},
  journal = {Int. J. Quant. Chem.},
  volume  = {114},
  pages   = {14},
  year    = {2014},
  DOI = {10.1002/qua.24521}
}

@article{Heyd:2006,
  author  = {J.~Heyd and G.E.~Scuseria and M.~Ernzerhof},
  title   = {{Erratum: “Hybrid functionals based on a screened Coulomb potential” [J. Chem. Phys. 118, 8207 (2003)]}},
  journal = {J. Chem. Phys.},
  volume  = {124},
  pages   = {219906},
  year    = {2006},
  DOI = {10.1063/1.2204597}
}

@article{Hohenberg:1964,
  author  = {P.~Hohenberg and W.~Kohn},
  title   = {Inhomogeneous electron gas},
  journal = {Phys. Rev.},
  volume  = {136},
  pages   = {B864},
  year    = {1964},
  DOI = {10.1103/PhysRev.136.B864}
}

@article{Kohn:1965,
  author  = {W.~Kohn and L.J.~Sham},
  title   = {Self-Consistent Equations Including Exchange and Correlation Effects},
  journal = {Phys. Rev.},
  volume  = {140},
  pages   = {A1133},
  year    = {1965},
  DOI = {10.1103/PhysRev.140.A1133}
}

@article{KirchnerHall:2021,
  author  = {Kirchner-Hall, N.E. and Zhao, W. and Xiong, Y. and Timrov, I. and Dabo, I.},
  title   = {{Extensive Benchmarking of DFT+$U$ Calculations for Predicting Band Gaps}},
  journal = {Appl. Sci.},
  volume  = {11},
  pages   = {2395},
  year    = {2021},
  doi = {10.3390/app11052395},
}

@article{Yu:2020,
  author  = {Yu, M. and Yang, S. and Wu, C. and Marom, N.},
  title   = {{Machine learning the Hubbard $U$ parameter in DFT+$U$ using Bayesian optimization}},
  journal = {npj Comput. Mater.},
  volume  = {6},
  pages   = {180},
  year    = {2020},
  DOI = {10.1038/s41524-020-00446-9}
}

@article{Timrov:2018,
  author  = {I.~Timrov and N.~Marzari and M.~Cococcioni},
  title   = {Hubbard parameters from density-functional perturbation theory},
  journal = {Phys. Rev. B},
  volume  = {98},
  pages   = {085127},
  year    = {2018},
  DOI = {10.1103/PhysRevB.98.085127}
}

@article{Timrov:2020b,
  author  = {I.~Timrov and F.~Aquilante and L.~Binci and M.~Cococcioni and N.~Marzari},
  title   = {{Pulay forces in density-functional theory with extended Hubbard functionals: from nonorthogonalized to orthogonalized manifolds}},
  journal = {Phys. Rev. B},
  volume  = {102},
  pages   = {235159},
  year    = {2020},
  DOI = {10.1103/PhysRevB.102.235159}
}

@article{Timrov:2021,
  author  = {I.~Timrov and N.~Marzari and M.~Cococcioni},
  title   = {{Self-consistent Hubbard parameters from density-functional perturbation theory in the ultrasoft and projector-augmented wave formulations}},
  journal = {Phys. Rev. B},
  volume  = {103},
  pages   = {045141},
  year    = {2021},
  DOI = {10.1103/PhysRevB.103.045141}
}

@article{Timrov:2022,
  author  = {I.~Timrov and N.~Marzari and M.~Cococcioni},
  title   = {{\texttt{HP} -- A code for the calculation of Hubbard parameters using density-functional perturbation theory}},
  journal = {Comput. Phys. Commun.},
  volume  = {279},
  pages   = {108455},
  year    = {2022},
  DOI = {10.24435/materialscloud:v6-zd}
}

@article{Timrov:2022b,
  author  = {I.~Timrov and F.~Aquilante and M.~Cococcioni and N.~Marzari},
  title   = {{Accurate Electronic Properties and Intercalation Voltages of Olivine-type Li-ion Cathode Materials from Extended Hubbard Functionals}},
  journal = {PRX Energy},
  volume  = {1},
  pages   = {033003},
  year    = {2022},
  DOI = {10.1103/PRXEnergy.1.033003}
}

@Article{Timrov:2023,
author = {Timrov, I. and Kotiuga, M. and Marzari, N.},
title = {{Unraveling the effects of inter-site Hubbard interactions in spinel Li-ion cathode materials}},
journal = {Phys. Chem. Chem. Phys.},
volume = {25},
pages = {9061},
year ={2023},
doi = {10.1039/d3cp00419h},
}

@article{Garrity:2014,
  author  = {Garrity, K.F. and Bennett, J.W. and Rabe, K.M. and Vanderbilt, D.},
  title = {Pseudopotentials for high-throughput DFT calculations},
  journal = {Comput. Mater. Sci.},
  volume  = {81},
  pages   = {446},
  year    = {2014},
  doi = {10.1016/j.commatsci.2013.08.053},
}

@article{Gautam:2018,
  author  = {Gautam, G.S. and Carter, E.},
  title   = {{Evaluating transition metal oxides within DFT-SCAN and SCAN+$U$ frameworks for solar thermochemical applications}},
  journal = {Phys. Rev. Materials},
  volume  = {2},
  pages   = {095401},
  year    = {2018},
  DOI = {10.1103/PhysRevMaterials.2.095401}
}

@article{Kaczkowski:2021,
  author  = {Kaczkowski, J. and Pugaczowa-Michalska, M. and P\'lowa\'s-Korus, I.},
  title   = {{Comparative density functional studies of pristine and doped bismuth ferrite polymorphs by GGA+$U$ and meta-GGA SCAN+$U$}},
  journal = {Phys. Chem. Chem. Phys.},
  volume  = {23},
  pages   = {8571},
  year    = {2021},
  DOI = {10.1039/D0CP06157C}
}

@book{Kubaschewski:1993,
  author    = {Kubaschewski, O. and Alcock, C.B. and Spencer, P.J.},
  title     = {{Materials thermochemistry}},
  edition   = {6th},
  publisher = {Pergamon Press: Elmsford},
  address   = {New York},
  year      = {1993},
  notes     = {Ch. 5, pp. 257-323}
}

@article{Giannozzi:2009,
  author  = {Giannozzi, P. and Baroni, S. and Bonini, N. and Calandra, M. and Car, R. and Cavazzoni, C. and Ceresoli, D. and Chiarotti, G.L. and Cococcioni, M. and Dabo, I. and Dal Corso, A. and De Gironcoli, S. and Fabris, S. and Fratesi, G. and Gebauer, R. and Gerstmann, U. and Gougoussis, C. and Kokalj, A. and Lazzeri, M. and Martin-Samos, L. and Marzari, N. and Mauri, F. and Mazzarello, R. and Paolini, S. and Pasquarello, A. and Paulatto, L. and Sbraccia, C. and Scandolo, S. and Sclauzero, G. and Seitsonen, A.P. and Smogunov, A. and Umari, P. and Wentzcovitch, R.M. },
  title   = {{Q}uantum {ESPRESSO}: {A} modular and open-source software project for quantum simulations of materials},
  journal = {J. Phys.: Condens. Matter.},
  year    = {2009},
  volume  = {21},
  pages   = {395502},
  DOI = {10.1088/0953-8984/21/39/395502}
}

@article{Giannozzi:2017,
  author  = {Giannozzi, P. and Andreussi, O. and Brumme, T. and Bunau, O. and
             Buongiorno~Nardelli, M. and Calandra, M. and Car, R. and Cavazzoni, C.
             and Ceresoli, D. and Cococcioni, M. and Colonna, N. and Carnimeo, I.
             and Dal~Corso, A. and de~Gironcoli, S. and Delugas, P. and  
             Di{S}tasio~{J}r., R.~A. and Ferretti, A. and Floris, A. and 
             Fratesi, G. and Fugallo, G. and Gebauer, R. and Gerstmann, U.
             and Giustino, F. and Gorni, T. and Jia, J. and Kawamura, M. and 
             Ko, H.-Y. and Kokalj, A. and K\"{u}\c{c}\"{u}kbenli, E. and
             Lazzeri, M. and Marsili, M. and Marzari, N. and Mauri, F. and 
             Nguyen, N.~L. and Nguyen, H.-V. and Otero-de-la-{R}osa, A.
             and Paulatto, L. and Ponc\'e, S. and Rocca, D. and Sabatini, R. 
             and Santra, B. and Schlipf, M. and Seitsonen, A.P. and 
             Smogunov, A. and Timrov, I. and Thonhauser, T. and Umari, P.
             and Vast, N. and Baroni, S.},
  title   = {{A}dvanced capabilities for materials modelling with {Q}uantum {ESPRESSO}},
  journal = {J. Phys.: Condens. Matter.},
  year    = {2017},
  volume  = {29},
  pages   = {465901},
  doi     = {10.1088/1361-648X/aa8f79}
}

@article{Giannozzi:2020,
  author  = {Giannozzi, P. and Baseggio, O. and Bonf\`a, P. and Brunato, D. and Car, R. and Carnimeo, I. and Cavazzoni, C. and de~Gironcoli, S. and Delugas, P. and Ferrari~Ruffino, F. and Ferretti, A. and Marzari, N. and Timrov, I. and Urru, A. and Baroni, S.},
  title   = {{Quantum ESPRESSO toward the exascale}},
  journal = {J. Chem. Phys.},
  volume  = {152},
  pages   = {154105},
  year    = {2020},
  doi     = {10.1063/5.0005082}
}

@article{Perdew:1981,
  author  = {J.P.~Perdew and A.~Zunger},
  title   = {Self-interaction correction to density-functional approximations for many-electron systems},
  journal = {Phys. Rev. B},
  volume  = {23},
  pages   = {5048},
  year    = {1981},
  DOI = {10.1103/PhysRevB.23.5048}
}

@article{Perdew:2008,
  author  = {J.P.~Perdew and A.~Ruzsinszky and G.I.~Csonka and O.A.~Vydrov and G.E.~Scuseria and L.A.~Constantin and X.~Zhou and K.~Burke},
  title   = {{Restoring the Density-Gradient Expansion for Exchange in Solids and Surfaces}},
  journal = {Phys. Rev. Lett.},
  volume  = {100},
  pages   = {136406},
  year    = {2008},
  notes   = {ibid. {\bf 102}, 039902 (2009)},
  DOI = {10.1103/PhysRevLett.100.136406}
}

@article{Lowdin:1950,
  author  = {L\"owdin, P.-O.},
  title   = {{On the Non‐Orthogonality Problem Connected with the Use of Atomic Wave Functions in the Theory of Molecules and Crystals}},
  journal = {J. Chem. Phys.},
  volume  = {18},
  pages   = {365},
  year    = {1950},
  DOI = {10.1063/1.1747632}
}

@article{Long:2020,
  author  = {Long, O.Y. and Gautam, G.S. and Carter, E.A.},
  title   = {{Evaluating optimal $U$ for $3d$ transition-metal oxides within the SCAN+$U$ framework}},
  journal = {Phys. Rev. Materials},
  volume  = {4},
  pages   = {045401},
  year    = {2020},
  DOI = {10.1103/PhysRevMaterials.4.045401}
}

@article{Liechtenstein:1995,
  author  = {A.I.~Liechtenstein and V.I.~Anisimov and J.~Zaanen},
  title   = {{Density-functional theory and strong interactions: Orbital ordering in Mott-Hubbard insulators}},
  journal = {Phys. Rev. B},
  volume  = {52},
  pages   = {R5467},
  year    = {1995},
  DOI = {10.1103/PhysRevB.52.R5467}
}

@Article{Kulik:2006,
author = {H.J.~Kulik and M.~Cococcioni and D.A.~Scherlis and N.~Marzari},
title = {{Density Functional Theory in Transition-Metal Chemistry: A Self-Consistent Hubbard U Approach}},
journal = {Phys. Rev. Lett.},
volume = {97},
pages = {103001},
year ={2006},
doi = {10.1103/PhysRevLett.97.103001},
}

@article{Kulik:2008,
  author  = {H.J.~Kulik and N.~Marzari},
  title   = {{A self-consistent Hubbard U density-functional theory approach to the addition-elimination reactions of hydrocarbons on bare FeO$^+$}},
  journal = {J. Chem. Phys.},
  volume  = {129},
  pages   = {134314},
  year    = {2008},
  DOI = {10.1063/1.2987444}
}

@article{Campo:2010,
  title     = {{Extended DFT+$U$+$V$ method with on-site and inter-site electronic interactions}},
  author    = {Campo Jr, Vivaldo Leiria and Cococcioni, Matteo},
  journal   = {J. Phys.: Condens. Matter},
  volume    = {22},
  number    = {5},
  pages     = {055602},
  year      = {2010},
  publisher = {IOP Publishing},
  DOI = {10.1088/0953-8984/22/5/055602}
}

@article{Sun:2015,
  author  = {J.~Sun and A.~Ruzsinszky and J.P.~Perdew},
  title   = {{Strongly Constrained and Appropriately Normed Semilocal Density Functional}},
  journal = {Phys. Rev. Lett.},
  volume  = {115},
  pages   = {036402},
  year    = {2015},
  DOI = {10.1103/PhysRevLett.115.036402}
}

@article{Bartok:2019,
  author  = {Bart\'ok, A.P. and Yates, J.R.},
  title   = {{Regularized SCAN functional}},
  journal = {J. Chem. Phys.},
  volume  = {150},
  pages   = {161101},
  year    = {2019},
  DOI = {10.1063/1.5094646}
}

@article{Furness:2020,
  author  = {Furness, J.W. and Kaplan, A.D. and Ning, J. and Perdew, J.P. and Sun, J.},
  title   = {{Accurate and Numerically Efficient r$^2$SCAN Meta-Generalized Gradient Approximation}},
  journal = {J. Phys. Chem. Lett.},
  volume  = {11},
  pages   = {8208},
  year    = {2020},
  DOI = {10.1021/acs.jpclett.0c02405}
}

@book{Fletcher:1987,
  author    = {Fletcher, R.},
  title     = {{Practical Methods of Optimization}},
  edition   = {2nd},
  publisher = {Wiley},
  address   = {Chichester},
  year      = {1987}
}

@article{Mahajan:2021,
  author  = {Mahajan, R. and Timrov, I. and Marzari, N. and Kashyap, A.},
  title   = {{Importance of intersite Hubbard interactions in $\beta$-MnO$_2$: A first-principles DFT+$U$+$V$ study}},
  journal = {Phys. Rev. Materials},
  volume  = {5},
  pages   = {104402},
  year    = {2021},
  doi = {10.1103/PhysRevMaterials.5.104402},
}

@Article{Mahajan:2022,
author = {Mahajan, R.  and Kashyap, A. and Timrov, I.},
title = {{Pivotal Role of Intersite Hubbard Interactions in Fe-Doped $\alpha$-MnO$_2$}},
journal = {J. Phys. Chem. C},
volume = {126},
pages = {14353},
year ={2022},
doi={https://doi.org/10.1021/acs.jpcc.2c04767},
}

@article{Mayer:2002,
  author  = {I.~Mayer},
  title   = {{On L\"owdin's method of symmetric orthogonalization}},
  journal = {Int. J. Quant. Chem.},
  volume  = {90},
  pages   = {63},
  year    = {2002},
  DOI = {10.1002/qua.981}
}

@article{MoriSanchez:2006,
  author  = {Mori-S\'anchez, P. and Cohen, A.J. and Yang, W.},
  title   = {{Many-electron self-interaction error in approximate density functionals}},
  journal = {J. Chem. Phys.},
  volume  = {125},
  pages   = {201102},
  year    = {2006},
  DOI = {10.1063/1.2403848}
}

@article{Wang:2006,
  author  = {Wang, L. and Maxisch, T. and Ceder, G.},
  title   = {{Oxidation energies of transition metal oxides within the GGA+U framework}},
  journal = {Phys. Rev. B},
  volume  = {73},
  pages   = {195107},
  year    = {2006},
  DOI = {10.1103/PhysRevB.73.195107}
}

@article{Warda:2026,
  author  = {Warda, K. and Macke, E. and Timrov, I. and Ciacchi, L.C. and Kowalski, P.M.},
  title   = {{Getting the manifold right: The crucial role of orbital resolution in DFT+$U$ for mixed $d$-$f$ electron compounds}},
  journal = {J. Chem. Theory Comput.},
  volume  = {22},
  pages   = {1016},
  year    = {2026},
  doi = {10.1021/acs.jctc.5c01406}
}

@article{dosSantos:2026,
  author  = {{dos Santos}, F.J. and Binci, L. and Menichetti, G. and Mahajan, R. and Marzari, N. and Timrov, I.},
  title   = {{Comparative study of magnetic exchange parameters and magnon dispersions in NiO and MnO from first principles}},
  journal = {Phys. Rev. B},
  volume  = {113},
  pages   = {024427},
  year    = {2026},
  doi = {10.1103/gtxm-6vtg}
}

@article{DalCorso:2014,
  author  = {{Dal Corso}, A.},
  title   = {{Pseudopotentials periodic table: From H to Pu}},
  journal = {Comput. Mater. Sci.},
  volume  = {95},
  pages   = {337},
  year    = {2014},
  doi = {10.1016/j.commatsci.2014.07.043},
}

@article{LeBacq:2004,
  author  = {{Le Bacq}, O. and Pasturel, A. and Bengone, O.},
  title   = {{Impact on electronic correlations on the structural stability, magnetism, and voltage of LiCoPO$_4$ battery}},
  journal = {Phys. Rev. B},
  volume  = {69},
  pages   = {245107},
  year    = {2004},
  DOI = {10.1103/PhysRevB.69.245107}
}

@article{Linscott:2018,
  title   = {{Role of spin in the calculation of Hubbard $U$ and Hund's $J$ parameters from first principles}},
  author  = {Linscott, E.B. and Cole, D.J. and Payne, M.C. and {O'Regan}, D.D.},
  journal = {Phys. Rev. B},
  volume  = {98},
  pages   = {235157},
  year    = {2018},
  doi = {10.1103/PhysRevB.98.235157},
}

@article{Chai:2024,
  title   = {{Minimum Tracking Linear Response Hubbard and Hund Corrected Density Functional Theory in CP2K}},
  author  = {Z. Chai and R. Si and M. Chen and G. Teobaldi and {O'Regan}, D.D. and L.-M. Liu},
  journal = {J. Chem. Theory Comput.},
  volume  = {20},
  pages   = {8984},
  year    = {2024},
  doi = {10.1021/acs.jctc.4c00964},
}

@article{MacEnulty:2024,
  title   = {{Facilities and practices for linear response Hubbard parameters $U$ and $J$ in Abinit}},
  author  = {L. {MacEnulty} and M. Giantomassi and B Amadon and G.-M. Rignanese and D.D. {O’Regan}},
  journal = {Electron. Struct.},
  volume  = {6},
  pages   = {037003},
  year    = {2024},
  doi = {10.1088/2516-1075/ad610f},
}

@article{Dudarev:1998,
  title     = {{Electron-energy-loss spectra and the structural stability of nickel oxide: An LSDA+$U$ study}},
  author    = {Dudarev, SL and Botton, GA and Savrasov, SY and Humphreys, CJ and Sutton, AP},
  journal   = {Phys. Rev. B},
  volume    = {57},
  number    = {3},
  pages     = {1505},
  year      = {1998},
  publisher = {APS},
  DOI = {10.1103/PhysRevB.57.1505}
}

@article{TancogneDejean:2020,
  author  = {Tancogne-Dejean, N. and Rubio, A.},
  title   = {{Parameter-free hybridlike functional based on an extended Hubbard model: DFT+U+V}},
  journal = {Phys. Rev. B},
  volume  = {102},
  pages   = {155117},
  year    = {2020},
  DOI = {10.1103/PhysRevB.102.155117}
}

@article{Tavadze:2021,
  author  = {Tavadze, P. and Boucher, R. and Avendano-Franco, G. and Kocan, K.X. and Singh, S. and Dovale-Farelo, V. and Ibarra-Hern\'{a}ndez, W. and Johnson, M.B. and Mebane, D.S. and Romero, A.H. },
  title   = {{Exploring DFT+$U$ parameter space with a Bayesian calibration assisted by Markov chain Monte Carlo sampling}},
  journal = {npj Comput. Mater.},
  volume  = {7},
  pages   = {182},
  year    = {2021},
  DOI = {10.1038/s41524-021-00651-0}
}

@article{Lee:2020,
  author  = {Lee, S.-H. and Son, Y.-W.},
  title   = {{First-principles approach with a pseudohybrid density functional for extended Hubbard interactions}},
  journal = {Phys. Rev. Research},
  volume  = {2},
  pages   = {043410},
  year    = {2020},
  DOI = {10.1103/PhysRevResearch.2.043410}
}

@article{Dederichs:1984,
  author  = {P.H.~Dederichs and S.~Bl\"ugel and R.~Zeller and H.~Akai},
  title   = {{Ground States of Constrained Systems: Application to Cerium Impurities}},
  journal = {Phys. Rev. Lett.},
  volume  = {53},
  pages   = {2512},
  year    = {1984},
  DOI = {10.1103/PhysRevLett.53.2512}
}

@article{Mcmahan:1988,
  author  = {A.K.~McMahan and R.M.~Martin and S.~Satpathy},
  title   = {{Calculated effective Hamiltonian for La$_2$CuO$_4$ and solution in the impurity Anderson approximation}},
  journal = {Phys. Rev. B},
  volume  = {38},
  pages   = {6650},
  year    = {1988},
  DOI = {10.1103/PhysRevB.38.6650}
}

@article{Gunnarsson:1989,
  author  = {O.~Gunnarsson and O.K.~Andersen and O.~Jepsen and J.~Zaanen},
  title   = {{Density-functional calculation of the parameters in the Anderson model: Application to Mn in CdTe}},
  journal = {Phys. Rev. B},
  volume  = {39},
  pages   = {1708},
  year    = {1989},
  DOI = {10.1103/PhysRevB.39.1708}
}

@article{Hybertsen:1989,
  author  = {M.S.~Hybertsen and M.~Schl\"uter and N.E.~Christensen},
  title   = {{Calculation of Coulomb-interaction parameters for La$_2$CuO$_4$ using a constrained-density-functional approach}},
  journal = {Phys. Rev. B},
  volume  = {39},
  pages   = {9028},
  year    = {1989},
  DOI = {10.1103/PhysRevB.39.9028}
}

@article{Gunnarsson:1990,
  author  = {O.~Gunnarsson},
  title   = {{Calculation of parameters in model Hamiltonians}},
  journal = {Phys. Rev. B},
  volume  = {41},
  pages   = {514},
  year    = {1990},
  DOI = {10.1103/PhysRevB.41.514}
}

@article{Pickett:1998,
  author  = {W.E.~Pickett and S.C.~Erwin and E.C.~Ethridge},
  title   = {{Reformulation of the LDA+$U$ method for a local-orbital basis}},
  journal = {Phys. Rev. B},
  volume  = {58},
  pages   = {1201},
  year    = {1998},
  DOI = {10.1103/PhysRevB.58.1201}
}

@article{Solovyev:2005,
  author  = {I.V.~Solovyev and M.~Imada},
  title   = {{Screening of Coulomb interactions in transition metals}},
  journal = {Phys. Rev. B},
  volume  = {71},
  pages   = {045103},
  year    = {2005},
  DOI = {10.1103/PhysRevB.71.045103}
}

@article{Nakamura:2006,
  author  = {K.~Nakamura and R.~Arita and Y.~Yoshimoto and S.~Tsuneyuki},
  title   = {{First-principles calculation of effective onsite Coulomb interactions of $3d$ transition metals: Constrained local density functional approach with maximally localized Wannier functions}},
  journal = {Phys. Rev. B},
  volume  = {74},
  pages   = {235113},
  year    = {2006},
  DOI = {10.1103/PhysRevB.74.235113}
}

@article{Shishkin:2016,
  author  = {M.~Shishkin and H.~Sato},
  title   = {{Self-consistent parametrization of DFT+$U$ framework using linear response approach: Application to evaluation of redox potentials of battery cathodes}},
  journal = {Phys. Rev. B},
  volume  = {93},
  pages   = {085135},
  year    = {2016},
  DOI = {10.1103/PhysRevB.93.085135}
}

@article{Yu:2014,
  author  = {K.~Yu and E.A.~Carter},
  title   = {{Comparing ab initio methods of obtaining effective U parameters for closed-shell materials}},
  journal = {J. Chem. Phys.},
  volume  = {140},
  pages   = {121105},
  year    = {2014},
  DOI = {10.1063/1.4869718}
}

@article{Mosey:2007,
  author  = {N.J.~Mosey and E.A.~Carter},
  title   = {{Ab initio evaluation of Coulomb and exchange parameters for DFT+U calculations}},
  journal = {Phys. Rev. B},
  volume  = {76},
  pages   = {155123},
  year    = {2007},
  DOI = {10.1103/PhysRevB.76.155123}
}

@article{Mosey:2008,
  author  = {N.J.~Mosey and P.~Liao and E.A.~Carter},
  title   = {{Rotationally invariant ab initio evaluation of Coulomb and exchange parameters for DFT+U calculations}},
  journal = {J. Chem. Phys.},
  volume  = {129},
  pages   = {014103},
  year    = {2008},
  DOI = {10.1063/1.2943142}
}

@article{Andriotis:2010,
  author  = {A.N.~Andriotis and R.M.~Sheetz and M.~Menon},
  title   = {{LSDA+$U$ method: A calculation of the $U$ values at the Hartree-Fock level of approximation}},
  journal = {Phys. Rev. B},
  volume  = {81},
  pages   = {245103},
  year    = {2010},
  DOI = {10.1103/PhysRevB.81.245103}
}

@article{Agapito:2015,
  author  = {Agapito, L.A. and Curtarolo, S. and Buongiorno Nardelli, M.},
  title   = {{Reformulation of DFT+$U$ as a Pseudohybrid Hubbard Density Functional for Accelerated Materials Discovery}},
  journal = {Phys. Rev. X},
  volume  = {5},
  pages   = {011006},
  year    = {2015},
  DOI = {10.1103/PhysRevX.5.011006}
}

@article{Springer:1998,
  author  = {M.~Springer and F.~Aryasetiawan},
  title   = {{Frequency-dependent screened interaction in Ni within the random-phase approximation}},
  journal = {Phys. Rev. B},
  volume  = {57},
  pages   = {4364},
  year    = {1998},
  DOI = {10.1103/PhysRevB.57.4364}
}

@article{Kotani:2000,
  author  = {T.~Kotani},
  title   = {{{\it Ab initio} random-phase-approximation calculation of the frequency-dependent effective interaction between 3d electrons: Ni, Fe, and MnO}},
  journal = {J. Phys.: Condens. Matter},
  volume  = {12},
  pages   = {2413},
  year    = {2000},
  DOI = {10.1088/0953-8984/12/11/307}
}

@article{Aryasetiawan:2004,
  author  = {F.~Aryasetiawan and M.~Imada and A.~Georges and G.~Kotliar and S.~Biermann and A.I.~Lichtenstein},
  title   = {{Frequency-dependent local interactions and low-energy effective models from electronic structure calculations}},
  journal = {Phys. Rev. B},
  volume  = {70},
  pages   = {195104},
  year    = {2004},
  DOI = {10.1103/PhysRevB.70.195104}
}

@article{Aryasetiawan:2006,
  author  = {F.~Aryasetiawan and K.~Karlsson and O.~Jepsen and U.~Sc\"onberger},
  title   = {{Calculations of Hubbard $U$ from first-principles}},
  journal = {Phys. Rev. B},
  volume  = {74},
  pages   = {125106},
  year    = {2006},
  DOI = {10.1103/PhysRevB.74.125106}
}

@article{Cococcioni:2005,
  author  = {M.~Cococcioni and S. de Gironcoli},
  title   = {{Linear response approach to the calculation of the effective interaction parameters in the {LDA+U} method}},
  journal = {Phys. Rev. B},
  volume  = {71},
  pages   = {035105},
  year    = {2005},
  DOI = {10.1103/PhysRevB.71.035105}
}

@article{Isaacs:2017,
  author  = {Isaacs, E.B. and Marianetti, C.A.},
  title   = {{Compositional phase stability of strongly correlated electron materials within DFT+$U$}},
  journal = {Phys. Rev. B},
  volume  = {95},
  pages   = {045141},
  year    = {2017},
  DOI = {10.1103/PhysRevB.95.045141

}
}

@article{Uhrin:2025,
  title={{Machine learning Hubbard parameters with equivariant neural networks}},
  author={Uhrin, M. and Zadoks, A. and Binci, L. and Marzari, N. and Timrov, I.},
  journal={npj Comput. Mater.},
  volume={11},
  number={},
  pages={19},
  year={2025},
  doi = {10.1038/s41524-024-01501-5},
}

@Article{Chang:2025,
author = {B.K. Chang and \textbf{I. Timrov} and J. Park and J.-J. Zhou and N. Marzari and M. Bernardi},
title = {{First-principles electron-phonon interactions and polarons in the parent cuprate La$_2$CuO$_4$}},
journal = {Phys. Rev. Research},
volume = {7},
pages = {L012073},
year ={2025},
doi = {10.1103/PhysRevResearch.7.L012073},
}

@misc{Hubparam:2023,
  note = {{The selection of a specific first-principles method to compute $U$ can be influenced by the intended applications and the properties of interest. Notably, cRPA is the preferred method for determining $U$ in the context of subsequent DFT plus dynamical mean field theory (DFT+DMFT) calculations~\cite{Georges:1996}.}}
}

@misc{MaterialsCloudArchive2026,
  author       = {Wooil Yang and Iurii Timrov and Francesco Aquilante and Young-Woo Son},
  title        = {{Comparing Hubbard parameters from linear-response theory and a Hartree-Fock-based approach}},
  publisher    = {Materials Cloud Archive},
  doi          = {10.24435/materialscloud:69-v6},
  url          = {https://archive.materialscloud.org/record/2026.80},
  howpublished = {Materials Cloud Archive \textbf{2026.80} (2026)}
}

@article{Liu:2023,
  author  = {B.-L. Liu and Y.-C. Wang and Y. Liu and Y.-J. Xu and X. Chen and H.-Z. Song and Y. Bi and H.-F. Liu and H.-F. Song},
  title   = {{Comparative study of first-principles approaches for effective Coulomb interaction strength $U_{eff}$ between localized $f$-electrons: Lanthanide metals as an example}},
  journal = {J. Chem. Phys.},
  volume  = {158},
  pages   = {084108},
  year    = {2023},
  doi = {10.1063/5.0137264},
}

@article{Binci:2023,
    author = {L. Binci and M. Kotiuga and I. Timrov and Nicola Marzari},
    title = {{Hybridization driving distortions and multiferroicity in rare-earth nickelates}},
    journal = {Phys. Rev. Research},
    volume = {5},
    pages = {033146},
    year = {2023},
    doi = {10.1103/PhysRevResearch.5.033146},
}

@Article{Binci:2025,
author = {Binci, L. and Marzari, N. and Timrov, I.},
title = {{Magnons from time-dependent density-functional perturbation theory and nonempirical Hubbard functionals}},
journal = {npj Comput. Mater.},
volume = {11},
pages = {100},
year = {2025},
doi = {10.1038/s41524-025-01570-0}
}

@article{Haddadi:2024,
    author = {F. Haddadi and E. Linscott and I. Timrov and N. Marzari and M. Gibertini},
    title = {{On-site and inter-site Hubbard corrections in magnetic monolayers: The case of FePS$_3$ and CrI$_3$}},
    journal = {Phys. Rev. Materials},
    volume = {8},
    pages = {014007},
    year = {2024},
    doi = {10.1103/PhysRevMaterials.8.014007},
}

@Article{Bonfa:2024,
author = {Bonf\'a, P. and Onuorah, I.J. and Lang, F. and I. Timrov and Monacelli, L. and Wang, C. and Sun, X. and Petracic, O. and Pizzi, G. and Marzari, N. and Blundell, S.J. and {De Renzi}, R.},
title = {{Magnetostriction-driven muon localisation in an antiferromagnetic oxide}},
journal = {Phys. Rev. Lett.},
volume = {132},
pages = {046701},
year ={2024},
doi={10.1103/PhysRevLett.132.046701},
}

@article{Gelin:2024,
  title={{Ternary oxides of $s-$ and $p-$block metals for photocatalytic solar-to-hydrogen conversion}},
  author={S. Gelin and N.E. {Kirchner-Hall} and R.R. Katzbaer and M.J. Theibault and Y. Xiong and W. Zhao and M.M. Khan and E. Andrewlavage and P. Orbe and S.M. Baksa and M. Cococcioni and I. Timrov and Q. Campbell and H. Abruna and R.E. Schaak and I. Dabo},
  journal={PRX Energy},
  volume={3},
  pages={013007},
  year={2024},
  doi = {10.1103/PRXEnergy.3.013007},
}

@article{Macke:2024,
  title={{Orbital-Resolved DFT+$U$ for Molecules and Solids}},
  author={Macke, E. and Timrov, I. and Marzari, N. and L. Colombi Ciacchi},
  journal={J. Chem. Theory Comput.},
  volume={20},
  number={},
  pages={4824},
  year={2024},
  doi = {10.1021/acs.jctc.3c01403},
}

@article{Yang:2021,
  title = {Ab initio study of lattice dynamics of group IV semiconductors using pseudohybrid functionals for extended Hubbard interactions},
  author = {Yang, Wooil and Jhi, Seung-Hoon and Lee, Sang-Hoon and Son, Young-Woo},
  journal = {Phys. Rev. B},
  volume = {104},
  issue = {10},
  pages = {104313},
  numpages = {12},
  year = {2021},
  month = {Sep},
  publisher = {American Physical Society},
  doi = {10.1103/PhysRevB.104.104313},
}

@article{Yang:2022,
doi = {10.1088/1361-648X/ac6c69},
year = {2022},
month = {may},
publisher = {IOP Publishing},
volume = {34},
number = {29},
pages = {295601},
author = {Wooil Yang and Bo Gyu Jang and Young-Woo Son and Seung-Hoon Jhi},
title = {Lattice dynamical properties of antiferromagnetic oxides calculated using self-consistent extended Hubbard functional method},
journal = {Journal of Physics: Condensed Matter},
}

@article{Jang:2023,
  title = {Intersite Coulomb Interactions in Charge-Ordered Systems},
  author = {Jang, Bo Gyu and Kim, Minjae and Lee, Sang-Hoon and Yang, Wooil and Jhi, Seung-Hoon and Son, Young-Woo},
  journal = {Phys. Rev. Lett.},
  volume = {130},
  issue = {13},
  pages = {136401},
  numpages = {7},
  year = {2023},
  month = {Mar},
  publisher = {American Physical Society},
  doi = {10.1103/PhysRevLett.130.136401},
  url = {https://link.aps.org/doi/10.1103/PhysRevLett.130.136401}
}

@article{Bersch:2008,
  title = {{Band offsets of ultrathin high-$\ensuremath{\kappa}$ oxide films with Si}},
  author = {Bersch, Eric and Rangan, Sylvie and Bartynski, Robert Allen and Garfunkel, Eric and Vescovo, Elio},
  journal = {Phys. Rev. B},
  volume = {78},
  issue = {8},
  pages = {085114},
  numpages = {10},
  year = {2008},
  month = {Aug},
  publisher = {American Physical Society},
  doi = {10.1103/PhysRevB.78.085114},
  url = {https://link.aps.org/doi/10.1103/PhysRevB.78.085114}
}

@article{Chen:1998,
    author = {Chen, Yefan and Bagnall, D. M. and Koh, Hang-jun and Park, Ki-tae and Hiraga, Kenji and Zhu, Ziqiang and Yao, Takafumi},
    title = {Plasma assisted molecular beam epitaxy of ZnO on $c$-plane sapphire: Growth and characterization},
    journal = {J. Appl. Phys.},
    volume = {84},
    number = {7},
    pages = {3912-3918},
    year = {1998},
    month = {10},
    issn = {0021-8979},
    doi = {10.1063/1.368595},
    url = {https://doi.org/10.1063/1.368595},
    eprint = {https://pubs.aip.org/aip/jap/article-pdf/84/7/3912/19186519/3912\_1\_online.pdf},
}

@article{Kwei:1993,
	author = {Kwei, G. H. and Lawson, A. C. and Billinge, S. J. L. and Cheong, S. W.},
	date = {1993/03/01},
	doi = {10.1021/j100112a043},
	isbn = {0022-3654},
	journal = {J. Phys. Chem.},
	month = {03},
	number = {10},
	pages = {2368--2377},
	publisher = {American Chemical Society},
	title = {Structures of the ferroelectric phases of barium titanate},
	type = {doi: 10.1021/j100112a043},
	url = {https://doi.org/10.1021/j100112a043},
	volume = {97},
	year = {1993},
}

@article{Wemple:1970,
  title = {{Polarization Fluctuations and the Optical-Absorption Edge in BaTi${\mathrm{O}}_{3}$}},
  author = {Wemple, S. H.},
  journal = {Phys. Rev. B},
  volume = {2},
  issue = {7},
  pages = {2679},
  numpages = {0},
  year = {1970},
  month = {Oct},
  publisher = {American Physical Society},
  doi = {10.1103/PhysRevB.2.2679},
  url = {https://link.aps.org/doi/10.1103/PhysRevB.2.2679}
}

@article{Tsunoda2019prm,
  title = {{Stabilization of small polarons in ${\mathrm{BaTiO}}_{3}$ by local distortions}},
  author = {Tsunoda, Naoki and Kumagai, Yu and Oba, Fumiyasu},
  journal = {Phys. Rev. Mater.},
  volume = {3},
  issue = {11},
  pages = {114602},
  numpages = {6},
  year = {2019},
  month = {Nov},
  publisher = {American Physical Society},
  doi = {10.1103/PhysRevMaterials.3.114602},
  url = {https://link.aps.org/doi/10.1103/PhysRevMaterials.3.114602}
}

@article{Gebreyesus:2023,
  title={{Understanding the role of Hubbard corrections in the rhombohedral phase of BaTiO$_3$}},
  author={G. Gebreyesus and L. Bastonero and M. Kotiuga and N. Marzari and I. Timrov},
  journal={Phys. Rev. B},
  volume={108},
  pages={235171},
  year={2023},
  doi = {10.1103/PhysRevB.108.235171},
}

@article{Choi:2025,
      title={{First-principles study of dielectric properties of ferroelectric perovskite oxides with on-site and inter-site Hubbard interactions}}, 
      author={Min Chul Choi and Wooil Yang and Young-Woo Son and Se Young Park},
      journal = {npj Comput. Mater.},
      volume = {11},
      pages = {221},
      year = {2025},
      doi = {10.1038/s41524-025-01711-5},
}

@article{Mang1995ssc,
title = {Band gaps, crystal-field splitting, spin-orbit coupling, and exciton binding energies in {ZnO} under hydrostatic pressure},
journal = {Solid State Commun.},
volume = {94},
number = {4},
pages = {251-254},
year = {1995},
issn = {0038-1098},
doi = {https://doi.org/10.1016/0038-1098(95)00054-2},
url = {https://www.sciencedirect.com/science/article/pii/0038109895000542},
author = {A. Mang and K. Reimann and St. Rübenacke},
}

@article{Reynolds1999prb,
  title = {Valence-band ordering in {ZnO}},
  author = {Reynolds, D. C. and Look, D. C. and Jogai, B. and Litton, C. W. and Cantwell, G. and Harsch, W. C.},
  journal = {Phys. Rev. B},
  volume = {60},
  issue = {4},
  pages = {2340--2344},
  numpages = {0},
  year = {1999},
  month = {Jul},
  publisher = {American Physical Society},
  doi = {10.1103/PhysRevB.60.2340},
  url = {https://link.aps.org/doi/10.1103/PhysRevB.60.2340}
}

@article{Dong2004prb,
  title = {Electronic structure of nanostructured {ZnO} from x-ray absorption and emission spectroscopy and the local density approximation},
  author = {Dong, C. L. and Persson, C. and Vayssieres, L. and Augustsson, A. and Schmitt, T. and Mattesini, M. and Ahuja, R. and Chang, C. L. and Guo, J.-H.},
  journal = {Phys. Rev. B},
  volume = {70},
  issue = {19},
  pages = {195325},
  numpages = {5},
  year = {2004},
  month = {Nov},
  publisher = {American Physical Society},
  doi = {10.1103/PhysRevB.70.195325},
  url = {https://link.aps.org/doi/10.1103/PhysRevB.70.195325}
}

@article{Kurmaev2008prb,
  title = {Oxygen x-ray emission and absorption spectra as a probe of the electronic structure of strongly correlated oxides},
  author = {Kurmaev, E. Z. and Wilks, R. G. and Moewes, A. and Finkelstein, L. D. and Shamin, S. N. and Kune\ifmmode \check{s}\else \v{s}\fi{}, J.},
  journal = {Phys. Rev. B},
  volume = {77},
  issue = {16},
  pages = {165127},
  numpages = {5},
  year = {2008},
  month = {Apr},
  publisher = {American Physical Society},
  doi = {10.1103/PhysRevB.77.165127},
  url = {https://link.aps.org/doi/10.1103/PhysRevB.77.165127}
}

@article{Elp15301991prb,
  title = {{Electronic structure of MnO}},
  author = {van Elp, J. and Potze, R. H. and Eskes, H. and Berger, R. and Sawatzky, G. A.},
  journal = {Phys. Rev. B},
  volume = {44},
  issue = {4},
  pages = {1530},
  numpages = {0},
  year = {1991},
  month = {Jul},
  publisher = {American Physical Society},
  doi = {10.1103/PhysRevB.44.1530},
  url = {https://link.aps.org/doi/10.1103/PhysRevB.44.1530}
}

@article{Cheetham1983prb,
  title = {{Magnetic ordering and exchange effects in the antiferromagnetic solid solutions ${\mathrm{Mn}}_{x}{\mathrm{Ni}}_{1\ensuremath{-}x}\mathrm{O}$}},
  author = {Cheetham, A. K. and Hope, D. A. O.},
  journal = {Phys. Rev. B},
  volume = {27},
  issue = {11},
  pages = {6964},
  numpages = {0},
  year = {1983},
  month = {Jun},
  publisher = {American Physical Society},
  doi = {10.1103/PhysRevB.27.6964},
  url = {https://link.aps.org/doi/10.1103/PhysRevB.27.6964}
}

@article{Fender1968JCP,
    author = {Fender, B. E. F. and Jacobson, A. J. and Wedgwood, F. A.},
    title = {{Covalency Parameters in MnO, $\alpha$‐MnS, and NiO}},
    journal = {J. Chem. Phys.},
    volume = {48},
    number = {3},
    pages = {990},
    year = {1968},
    month = {02},
    issn = {0021-9606},
    doi = {10.1063/1.1668855},
    url = {https://doi.org/10.1063/1.1668855},
}

@article{Bowen1975JSSC,
title = {{Electrical and optical properties of FeO}},
journal = {Journal of Solid State Chemistry},
volume = {12},
number = {3},
pages = {355},
year = {1975},
issn = {0022-4596},
doi = {https://doi.org/10.1016/0022-4596(75)90340-0},
url = {https://www.sciencedirect.com/science/article/pii/0022459675903400},
author = {H.K. Bowen and D. Adler and B.H. Auker},
}

@article{Roth1958prb,
  title = {{Magnetic Structures of MnO, FeO, CoO, and NiO}},
  author = {Roth, W. L.},
  journal = {Phys. Rev.},
  volume = {110},
  issue = {6},
  pages = {1333},
  numpages = {0},
  year = {1958},
  month = {Jun},
  publisher = {American Physical Society},
  doi = {10.1103/PhysRev.110.1333},
  url = {https://link.aps.org/doi/10.1103/PhysRev.110.1333}
}

@article{Battle1979JPCSSP,
  title={{The magnetic structure of non-stoichiometric ferrous oxide}},
  author={Battle, PD and Cheetham, AK},
  journal={J. Phys. C: Solid State Phys.},
  volume={12},
  number={2},
  pages={337},
  year={1979},
  doi = {10.1088/0022-3719/12/2/021},
}

@article{Elp60901991prb,
  title = {{Electronic structure of CoO, Li-doped CoO, and ${\mathrm{LiCoO}}_{2}$}},
  author = {van Elp, J. and Wieland, J. L. and Eskes, H. and Kuiper, P. and Sawatzky, G. A. and de Groot, F. M. F. and Turner, T. S.},
  journal = {Phys. Rev. B},
  volume = {44},
  issue = {12},
  pages = {6090},
  numpages = {0},
  year = {1991},
  month = {Sep},
  publisher = {American Physical Society},
  doi = {10.1103/PhysRevB.44.6090},
  url = {https://link.aps.org/doi/10.1103/PhysRevB.44.6090}
}

@article{Khan1970prb,
  title = {{Magnetic Form Factor of ${\mathrm{Co}}^{++}$ Ion in Cobaltous Oxide}},
  author = {Khan, D. C. and Erickson, R. A.},
  journal = {Phys. Rev. B},
  volume = {1},
  issue = {5},
  pages = {2243},
  numpages = {0},
  year = {1970},
  month = {Mar},
  publisher = {American Physical Society},
  doi = {10.1103/PhysRevB.1.2243},
  url = {https://link.aps.org/doi/10.1103/PhysRevB.1.2243}
}

@article{Jauch2001prb,
  title = {{Crystallographic symmetry and magnetic structure of CoO}},
  author = {Jauch, W. and Reehuis, M. and Bleif, H. J. and Kubanek, F. and Pattison, P.},
  journal = {Phys. Rev. B},
  volume = {64},
  issue = {5},
  pages = {052102},
  numpages = {3},
  year = {2001},
  month = {Jul},
  publisher = {American Physical Society},
  doi = {10.1103/PhysRevB.64.052102},
  url = {https://link.aps.org/doi/10.1103/PhysRevB.64.052102}
}

@article{Sawatzky1984prl,
  title = {{Magnitude and Origin of the Band Gap in NiO}},
  author = {Sawatzky, G. A. and Allen, J. W.},
  journal = {Phys. Rev. Lett.},
  volume = {53},
  issue = {24},
  pages = {2339},
  numpages = {0},
  year = {1984},
  month = {Dec},
  publisher = {American Physical Society},
  doi = {10.1103/PhysRevLett.53.2339},
  url = {https://link.aps.org/doi/10.1103/PhysRevLett.53.2339}
}

@article{Carey1991jmr,
	author = {Carey, M. J. and Spada, F. E. and Berkowitz, A. E. and Cao, W. and Thomas, G.},
	journal = {Journal of Materials Research},
	number = {12},
	pages = {2680--2687},
	title = {Preparation and structural characterization of sputtered CoO, NiO, and Ni0.5Co0.5O thin epitaxial films},
	volume = {6},
	year = {1991}}

@article{Carta:2025,
	author = {A. Carta and I. Timrov and S. Beck and C. Ederer},
    title = {{Bridging constrained random-phase approximation and linear response theory for computing Hubbard parameters}},
	journal = {arXiv:2505.03698},
	number = {},
	pages = {},
	volume = {},
	year = {2025},
    url = {https://arxiv.org/abs/2505.03698},
}

@article{Smith1965ActaC,
author = "Smith, D. K. and Newkirk, W.",
title = "{The crystal structure of baddeleyite (monoclinic ZrO${\sb 2}$) and its relation to the polymorphism of ZrO${\sb 2}$}",
journal = "Acta Crystallographica",
year = "1965",
volume = "18",
number = "6",
pages = "983--991",
month = "Jun",
doi = {10.1107/S0365110X65002402},
url = {https://doi.org/10.1107/S0365110X65002402},
}

@Article{Bastonero:2025,
author = {L. Bastonero and C. Malica and E. Macke and M. Bercx and S. Huber, I. Timrov and N. Marzari},
title = {{First-principles Hubbard parameters with automated and reproducible workﬂows}},
journal = {npj Comput. Mater.},
volume = {11},
pages = {183},
year ={2025},
doi={10.1038/s41524-025-01685-4},
}

@article{Moore:2024,
  title = {{High-throughput determination of Hubbard $U$ and Hund $J$ values for transition metal oxides via the linear response formalism}},
  author = {Moore, Guy C. and Horton, Matthew K. and Linscott, Edward and Ganose, Alexander M. and Siron, Martin and O'Regan, David D. and Persson, Kristin A.},
  journal = {Phys. Rev. Mater.},
  volume = {8},
  issue = {1},
  pages = {014409},
  numpages = {17},
  year = {2024},
  month = {Jan},
  publisher = {American Physical Society},
  doi = {10.1103/PhysRevMaterials.8.014409},
}

@article{Paudel2008prb,
  title = {{First-principles calculation of the O vacancy in ZnO: A self-consistent gap-corrected approach}},
  author = {Paudel, Tula R. and Lambrecht, Walter R. L.},
  journal = {Phys. Rev. B},
  volume = {77},
  issue = {20},
  pages = {205202},
  numpages = {9},
  year = {2008},
  month = {May},
  publisher = {American Physical Society},
  doi = {10.1103/PhysRevB.77.205202},
  url = {https://link.aps.org/doi/10.1103/PhysRevB.77.205202}
}

@article{Lambert2023prr,
  title = {Use of $\mathrm{DFT}+\mathrm{U}+\mathrm{J}$ with linear response parameters to predict non-magnetic oxide band gaps with hybrid-functional accuracy},
  author = {Lambert, D. S. and O'Regan, D. D.},
  journal = {Phys. Rev. Res.},
  volume = {5},
  issue = {1},
  pages = {013160},
  numpages = {25},
  year = {2023},
  month = {Mar},
  publisher = {American Physical Society},
  doi = {10.1103/PhysRevResearch.5.013160},
  url = {https://link.aps.org/doi/10.1103/PhysRevResearch.5.013160}
}

@article{Kulik2010jcp,
    author = {Kulik, Heather J. and Marzari, Nicola},
    title = {{Systematic study of first-row transition-metal diatomic molecules: A self-consistent DFT+U approach}},
    journal = {J. Chem. Phys.},
    volume = {133},
    number = {11},
    pages = {114103},
    year = {2010},
    month = {09},
    issn = {0021-9606},
    doi = {10.1063/1.3489110},
}

@article{Wang2016jcp,
    author = {Wang, Yue-Chao and Chen, Ze-Hua and Jiang, Hong},
    title = {{The local projection in the density functional theory plus $U$ approach: A critical assessment}},
    journal = {The Journal of Chemical Physics},
    volume = {144},
    number = {14},
    pages = {144106},
    year = {2016},
    month = {04},
    issn = {0021-9606},
    doi = {10.1063/1.4945608},
    url = {https://doi.org/10.1063/1.4945608},
}

@article{Orhan:2020,
  title = {{First-principles Hubbard $U$ and Hund's $J$ corrected approximate density functional theory predicts an accurate fundamental gap in rutile and anatase ${\mathrm{TiO}}_{2}$}},
  author = {Orhan, Okan K. and O'Regan, David D.},
  journal = {Phys. Rev. B},
  volume = {101},
  pages = {245137},
  year = {2020},
  doi = {10.1103/PhysRevB.101.245137},
}

@article{Drabkin:1968,
  author={I.A. Drabkin and R.N. Emel’yanova and R. N. Iskenderov and Ya. M. Ksendzov},
  journal={Fiz. Tverd. Tela (Leningrad)},
  volume={10},
  pages={3082},
  year={1968},
}

@article{Gvishi:1972,
title = {{Hall mobility and defect structure in undoped and Cr or Ti-doped CoO at high temperature}},
journal = {J. Phys. Chem. Solids},
volume = {33},
number = {4},
pages = {893},
year = {1972},
issn = {0022-3697},
doi = {https://doi.org/10.1016/S0022-3697(72)80105-7},
url = {https://www.sciencedirect.com/science/article/pii/S0022369772801057},
author = {M. Gvishi and D.S. Tannhauser},
}

@article{ksendzov:1965,
  title={},
  author={Ksendzov, Y. M. and Drabkin, I. A.},
  journal={Fiz. Tverd. Tela (Leningrad)},
  volume={7},
  pages={1884},
  year={1965}
}

\clearpage
\onecolumngrid
\section*{Supplemental Information}

\setcounter{section}{0}
\setcounter{subsection}{0}
\setcounter{figure}{0}
\setcounter{table}{0}
\setcounter{equation}{0}
\renewcommand{\thesection}{S\arabic{section}}
\renewcommand{\thesubsection}{\thesection.\arabic{subsection}}
\renewcommand{\thefigure}{S\arabic{figure}}
\renewcommand{\thetable}{S\arabic{table}}
\renewcommand{\theequation}{S\arabic{equation}}

\providecommand{\theHsection}{}
\providecommand{\theHsubsection}{}
\providecommand{\theHfigure}{}
\providecommand{\theHtable}{}
\providecommand{\theHequation}{}
\renewcommand{\theHsection}{SI.\arabic{section}}
\renewcommand{\theHsubsection}{SI.\arabic{section}.\arabic{subsection}}
\renewcommand{\theHfigure}{SI.\arabic{figure}}
\renewcommand{\theHtable}{SI.\arabic{table}}
\renewcommand{\theHequation}{SI.\arabic{equation}}

\section{Self-consistent protocol with structural relaxation}

\begin{figure}[htbp]
    \centering
    \includegraphics[width=0.38\textwidth]{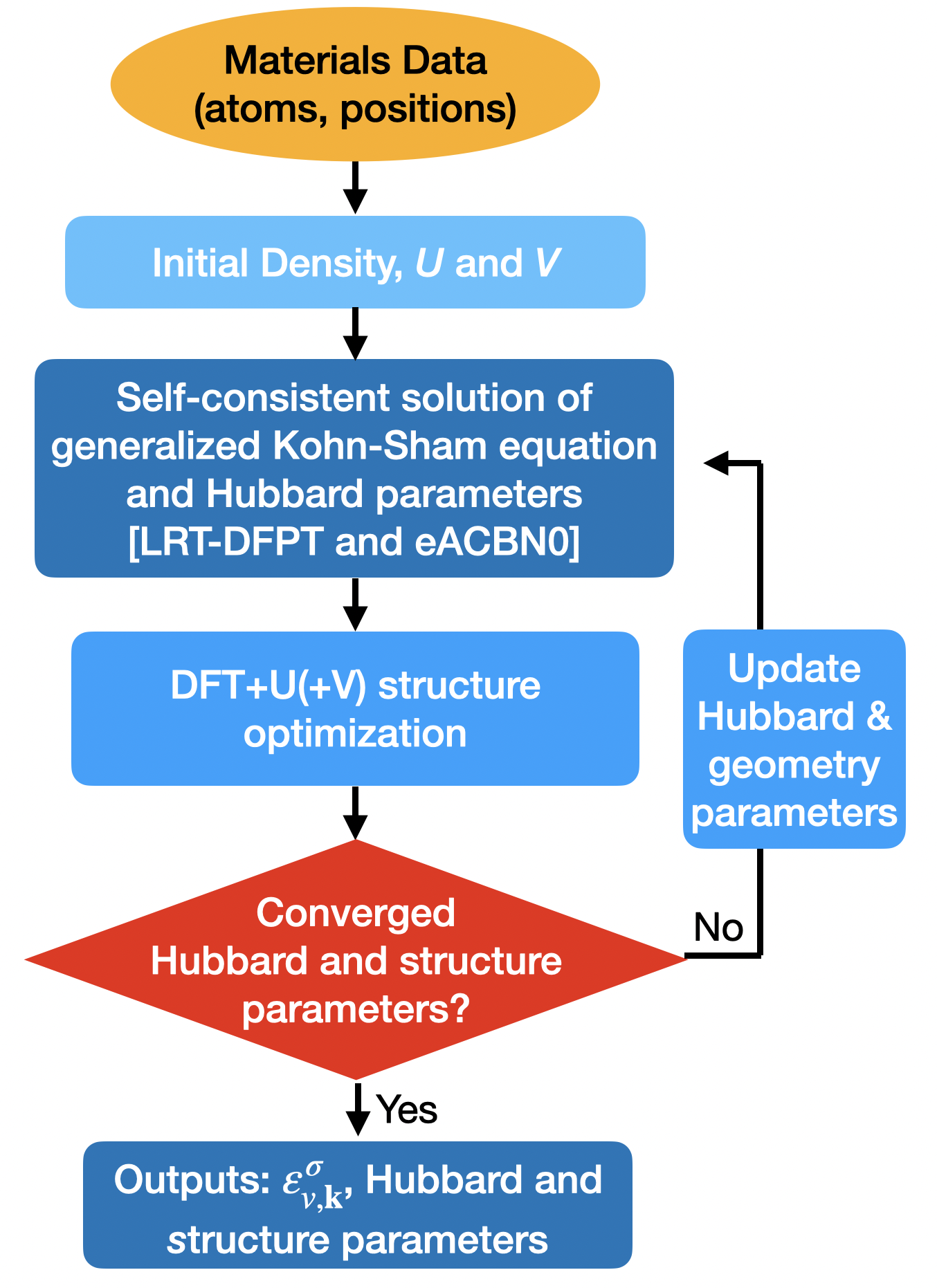}
    \caption{Protocol for computing Hubbard parameters within the ``structurally self-consistent'' scheme using either LRT (DFPT)~\cite{Timrov:2021} or eACBN0~\cite{Lee:2020}.
    Total energy, $U$, $V$, and the structural parameters are iteratively updated until each meets the chosen convergence threshold.}
    \label{fig:self-consistent}
\end{figure}

The computational workflow for the ``structurally self-consistent'' scheme, i.e. combining Hubbard-parameter calculations (via LRT or eACBN0) with structural relaxations, is illustrated in Fig.~\ref{fig:self-consistent}. The procedure begins with an initial crystal structure and an initial guess for the Hubbard parameters, which may be set to zero. A self-consistent DFT+$U$(+$V$) calculation is then performed to obtain the electronic ground state. In the case of eACBN0, the Hubbard parameters $U$ and $V$ are additionally computed as described in the main text. For clarity, we emphasize that LRT and eACBN0 are executed as two independent workflows.

Next, the crystal structure is optimized using DFT+$U$(+$V$), including Hubbard contributions to forces and stresses~\cite{Timrov:2020b,Yang:2021}, and employing the $U$ and $V$ values determined in the preceding step. If the updated Hubbard parameters or the relaxed geometry deviate from their previous values by more than user-defined thresholds, the cycle is repeated. This iterative loop continues until both the Hubbard parameters and the crystal structure are converged. The resulting self-consistent $U$ and $V$ values are then used in production calculations.

This protocol drives the system toward an energy extremum where the electronic and structural degrees of freedom are mutually consistent. Moreover, the same workflow can be applied to DFT+$U$ alone by simply setting $V=0$. The LRT (DFPT) workflow is fully automated in AiiDA-Hubbard, which greatly streamlines the entire procedure~\cite{Bastonero:2025}.

\newpage
\clearpage

\section{Self-consistent Hubbard parameters including structural relaxation}

\subsection*{MnO, FeO, CoO, and NiO}

\begin{figure}[htbp]
\centering
\includegraphics[width=0.8\textwidth]{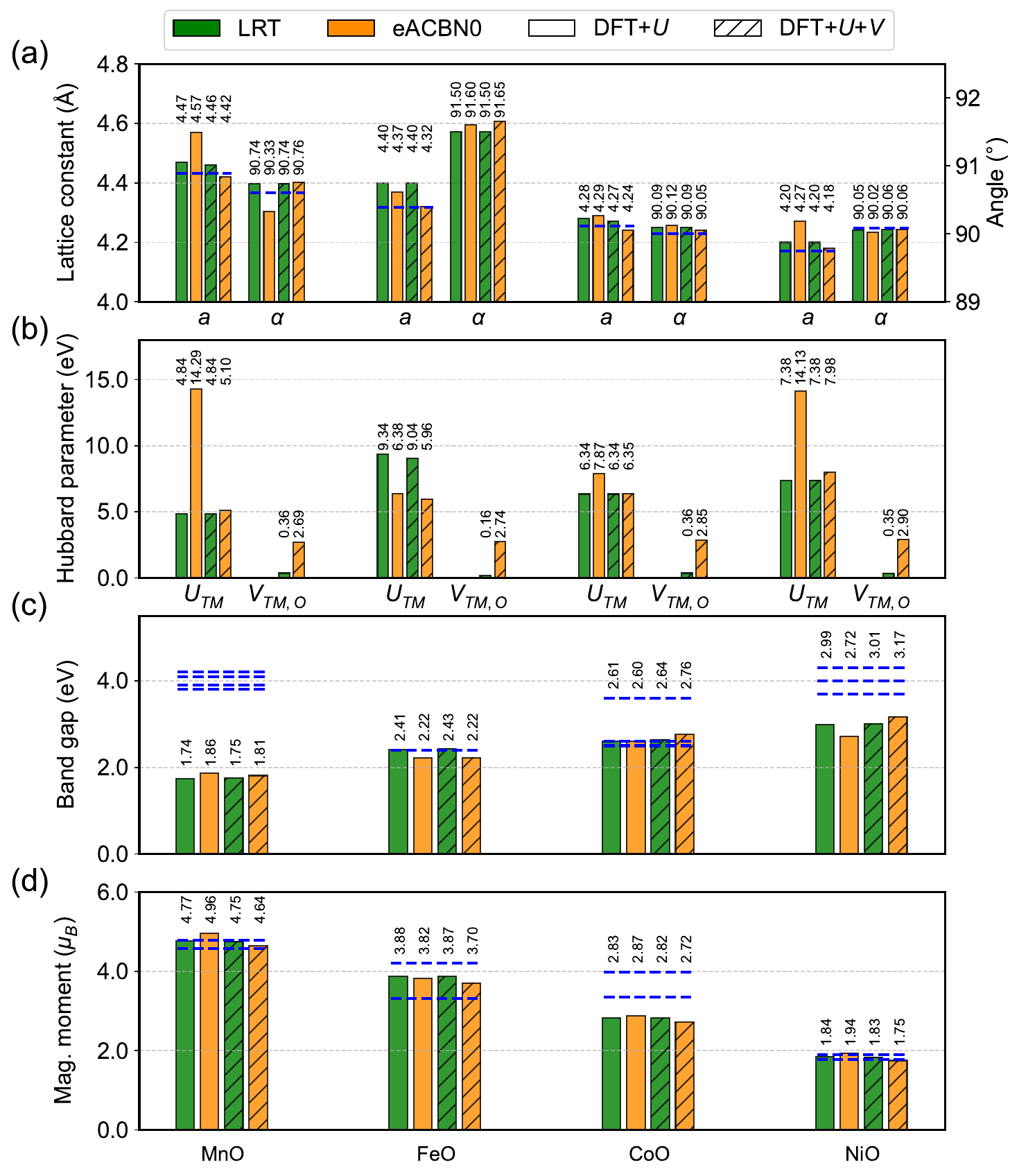}
\caption{DFT+$U$ and DFT+$U$+$V$ ``structurally self-consistent'' calculations for TMOs without the oxygen Hubbard correction ($U_\textrm{O}$). 
(a) Lattice constant $a$ and rhombohedral angle $\alpha$, 
(b) on-site Hubbard parameter $U_{\textrm{TM}}$ for the TM-$3d$ states and inter-site parameter $V_{\textrm{TM,O}}$ between TM-$3d$ and O-$2p$ states, 
(c)~band gap, and 
(d) magnetic moment. 
LRT and eACBN0 indicate the method used to obtain the Hubbard parameters. 
Experimental lattice constants and rhombohedral angles for MnO~\cite{Cheetham1983prb}, FeO~\cite{Battle1979JPCSSP}, CoO~\cite{Carey1991jmr}, and NiO~\cite{Cheetham1983prb} are shown as dashed blue lines in panel (a); for FeO, the values correspond to Fe$_{0.943}$O. 
Experimental band gaps and magnetic moments for MnO~\cite{Kurmaev2008prb,Elp15301991prb,Drabkin:1968,Cheetham1983prb,Fender1968JCP}, FeO~\cite{Bowen1975JSSC,Roth1958prb,Battle1979JPCSSP}, CoO~\cite{Kurmaev2008prb,Elp60901991prb,Gvishi:1972,Khan1970prb,Jauch2001prb}, and NiO~\cite{Kurmaev2008prb,Sawatzky1984prl,ksendzov:1965,Cheetham1983prb,Fender1968JCP} are indicated by dashed blue lines in panels (c) and (d), respectively.
}
\label{Fig:self_TMO_wo_O}
\end{figure}

\begin{figure}[htbp]
\centering
\includegraphics[width=0.8\textwidth]{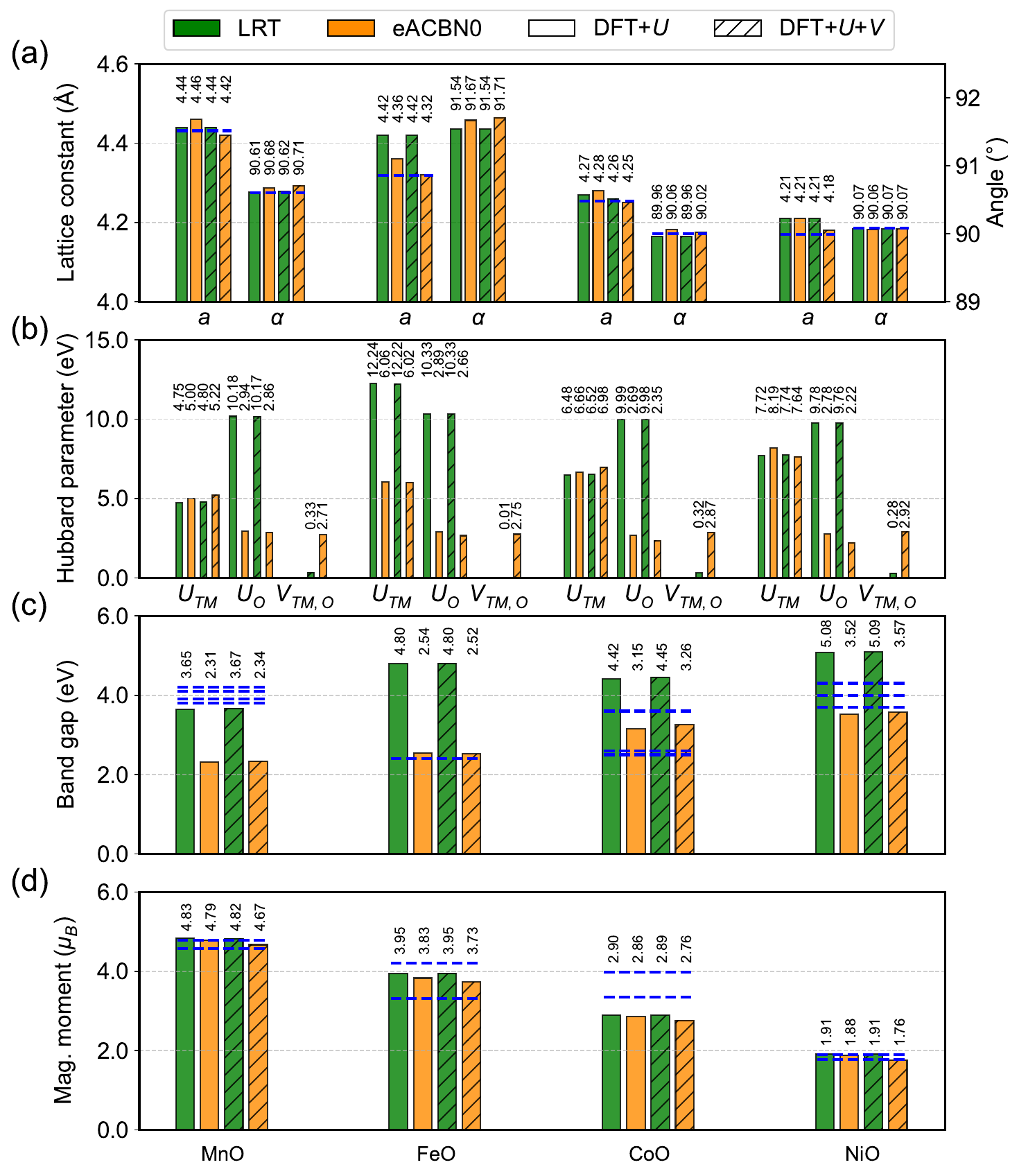}
\caption{DFT+$U$ and DFT+$U$+$V$ ``structurally self-consistent'' calculations for TMOs with the oxygen Hubbard correction ($U_\textrm{O}$). (a) Lattice constant $a$ and rhombohedral angle $\alpha$, 
(b) on-site Hubbard parameters $U_{\textrm{TM}}$ and $U_{\textrm{O}}$ for the TM-$3d$ states and the O-$2p$ states, respectively, and inter-site parameter $V_{\textrm{TM,O}}$ between TM-$3d$ and O-$2p$ states, 
(c)~band gap, and 
(d) magnetic moment. 
LRT and eACBN0 indicate the method used to obtain the Hubbard parameters. 
Experimental lattice constants and rhombohedral angles for MnO~\cite{Cheetham1983prb}, FeO~\cite{Battle1979JPCSSP}, CoO~\cite{Carey1991jmr}, and NiO~\cite{Cheetham1983prb} are shown as dashed blue lines in panel (a); for FeO, the values correspond to Fe$_{0.943}$O. 
Experimental band gaps and magnetic moments for MnO~\cite{Kurmaev2008prb,Elp15301991prb,Drabkin:1968,Cheetham1983prb,Fender1968JCP}, FeO~\cite{Bowen1975JSSC,Roth1958prb,Battle1979JPCSSP}, CoO~\cite{Kurmaev2008prb,Elp60901991prb,Gvishi:1972,Khan1970prb,Jauch2001prb}, and NiO~\cite{Kurmaev2008prb,Sawatzky1984prl,ksendzov:1965,Cheetham1983prb,Fender1968JCP} are indicated by dashed blue lines in panels (c) and (d), respectively.}
\label{Fig:self_TMO_with_O}
\end{figure}

For the four transition-metal oxides (TMOs) considered here, the relaxed geometries obtained with self-consistent Hubbard parameters show very good agreement with available experimental data, as illustrated in Figs.~\ref{Fig:self_TMO_wo_O} and~\ref{Fig:self_TMO_with_O}. These structural results are only weakly affected by the inclusion of the on-site Hubbard correction for oxygen ($U_O$), as evidenced by the close similarity between Fig.~\ref{Fig:self_TMO_wo_O}(a) and Fig.~\ref{Fig:self_TMO_with_O}(a).

The Hubbard parameters, band gaps, and magnetic moments obtained from the ``structurally self-consistent'' scheme (see Fig.~\ref{fig:self-consistent}) follow the same qualitative trends reported in the main text for calculations performed without structural relaxations (Figs.~2 and 3). Therefore, we conclude that full structural relaxations combined with self-consistent $U$ and $V$ values reliably reproduce the experimental geometries of MnO, CoO, and NiO, and that the associated electronic properties remain largely unchanged upon including structural relaxations.

\subsection*{ZrO$_2$ and BaTiO$_3$}

\begin{figure}[htbp]
\centering
\includegraphics[width=0.8\textwidth]{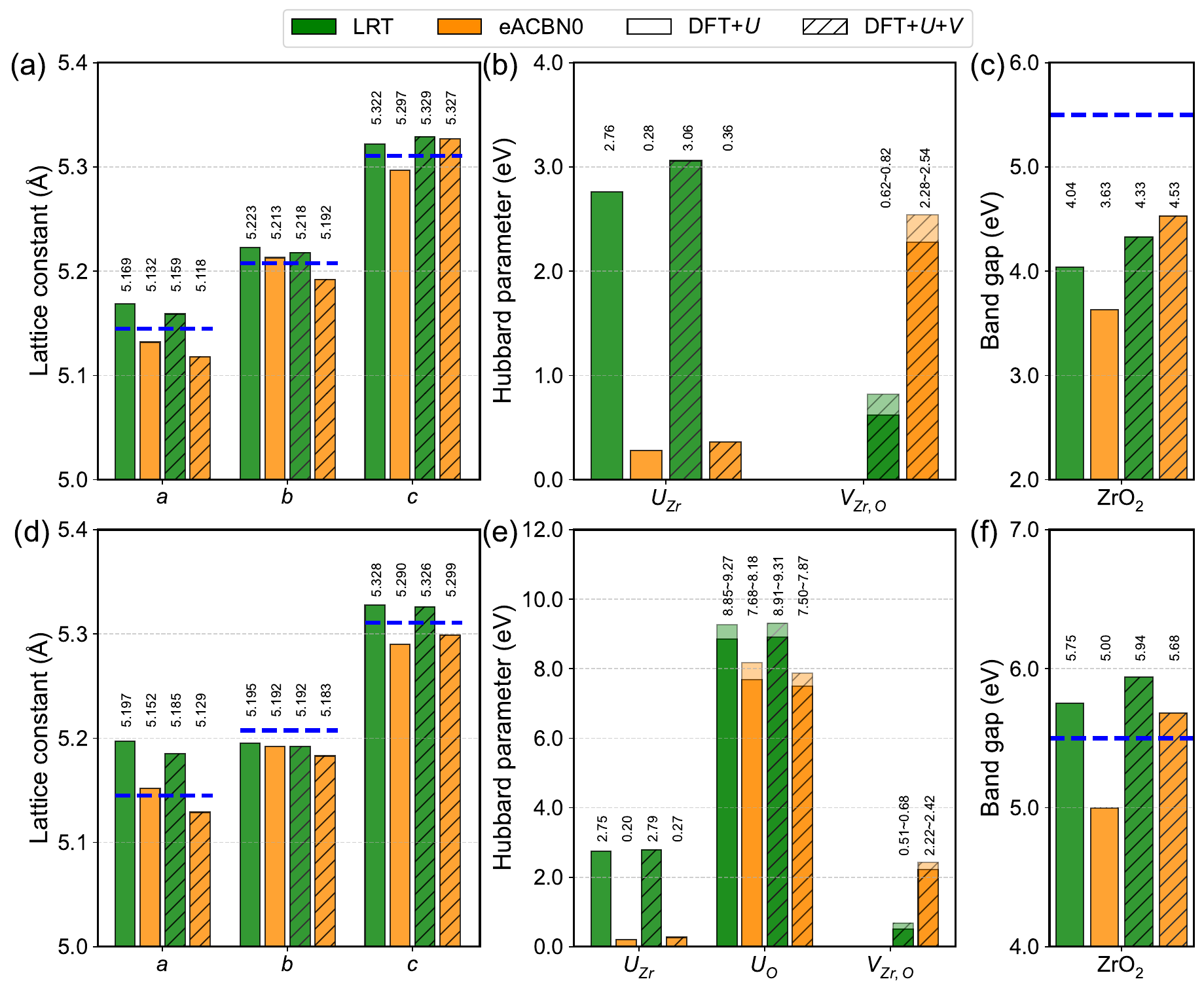}
\caption{DFT+$U$ and DFT+$U$+$V$ ``structurally self-consistent'' calculations for ZrO$_2$ without $U_\textrm{O}$ [panels (a)--(c)] and with $U_\textrm{O}$ [panels (d)--(f)]. 
Panels (a) and (d) show the lattice constants $a$, $b$, and $c$, with experimental values~\cite{Smith1965ActaC} indicated by dashed blue lines. 
Panels (b) and (e) display the on-site Hubbard parameters $U_{\textrm{Zr}}$ and $U_{\textrm{O}}$ for Zr-$4d$ and O-$2p$ states, respectively, as well as the inter-site parameter $V_{\textrm{Zr,O}}$ between Zr-$4d$ and O-$2p$ states. 
Panels (c) and (f) show the corresponding band gaps, with the experimental band gap~\cite{Bersch:2008} marked by a dashed blue line.
}
\label{Fig:self_ZrO2}
\end{figure}

Figure~\ref{Fig:self_ZrO2} summarizes the results for the nominally $d^0$ system ZrO$_2$ obtained using the ``structurally self-consistent'' scheme (see Fig.~\ref{fig:self-consistent}). The fully relaxed structural parameters show good agreement with experimental data, as illustrated in Figs.~\ref{Fig:self_ZrO2}(a) and (d). Consistently with results from the ``self-consistent'' scheme (no DFT+$U$+$V$ structural relaxation) presented in Figs.~4 and 5 of the main text, the LRT approach yields a finite $U_{\textrm{Zr}}$ value due to the response of small but non-zero $d$ occupations, whereas ACBN0 gives a vanishing value. The largest contribution to the band-gap correction arises from $U_\textrm{O}$ and $V_{\textrm{Zr,O}}$ (particularly within the ACBN0 scheme) which brings the calculated band gap closer to the experimental value. In contrast, the lattice constant is influenced mainly by the $U_{\textrm{Zr}}$ correction, with only minor sensitivity to $U_\textrm{O}$ and $V_{\textrm{Zr,O}}$.

\begin{figure}[htbp]
\centering
\includegraphics[width=0.8\textwidth]{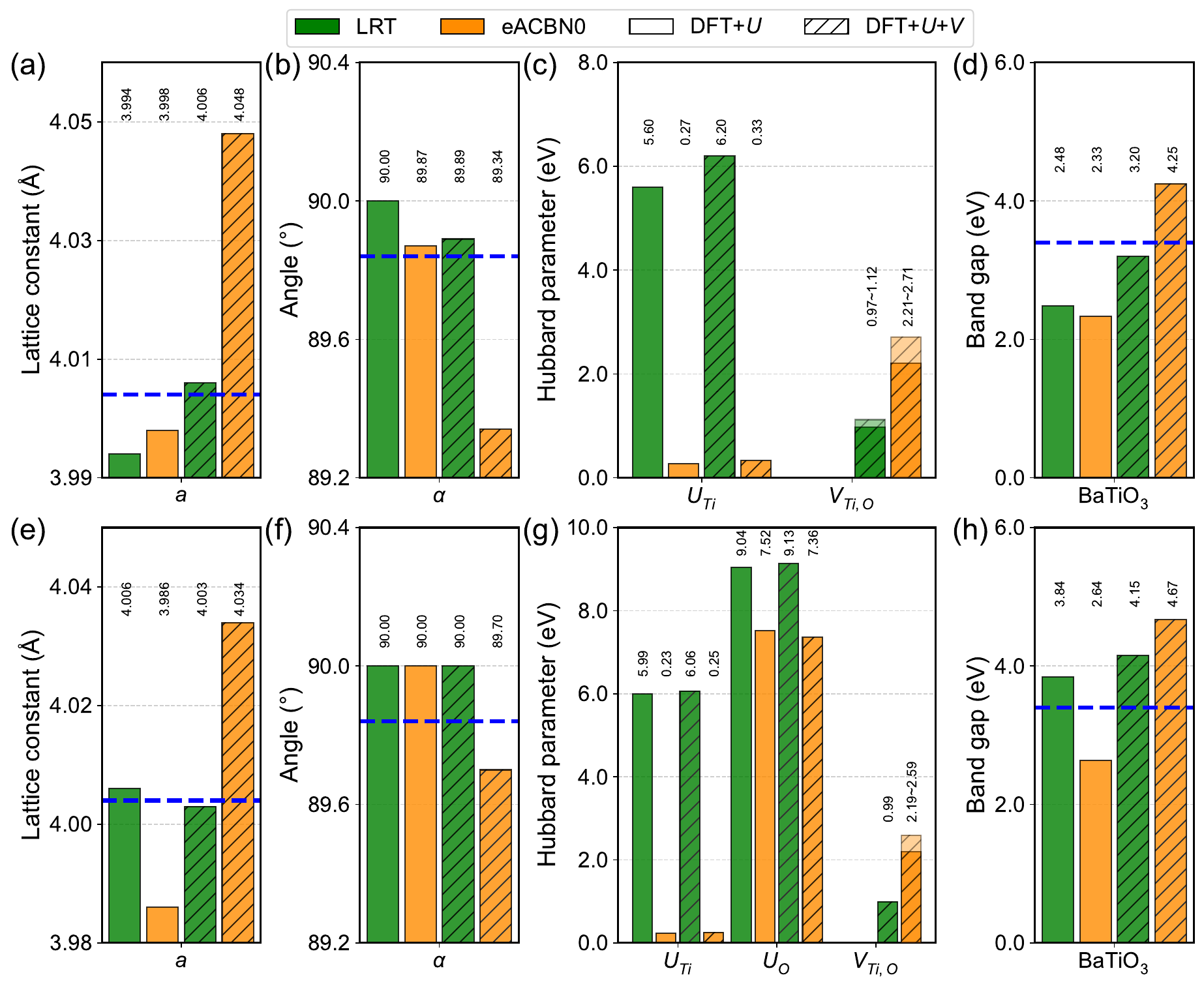}
\caption{DFT+$U$ and DFT+$U$+$V$ ``structurally self-consistent'' calculations for BaTiO$_3$ without $U_\textrm{O}$ [panels (a)--(d)] and with $U_\textrm{O}$ [panels (e)--(h)]. 
Panels (a) and (e) show the lattice constant $a$, while panels (b) and (f) display the rhombohedral angle $\alpha$, with experimental values~\cite{Kwei:1993} indicated by dashed blue lines. 
Panels (c) and (g) present the on-site Hubbard parameters $U_{\textrm{Ti}}$ and $U_{\textrm{O}}$ for Ti-$3d$ and O-$2p$ states, respectively, as well as the inter-site parameter $V_{\textrm{Ti,O}}$ between Ti-$3d$ and O-$2p$ states. 
Panels (d) and (h) show the corresponding band gaps, with the experimental band gap~\cite{Wemple:1970} marked by a dashed blue line.
}
\label{Fig:self_BaTiO3}
\end{figure}

Next, we consider BaTiO$_3$, which, like ZrO$_2$, has a nominal $d^0$ configuration. The results, shown in Fig.~\ref{Fig:self_BaTiO3}, follow trends consistent with those discussed in the main text. For the structural properties, when the on-site oxygen Hubbard correction ($U_\textrm{O}$) is excluded, LRT-based DFT+$U$+$V$ and eACBN0-based DFT+$U$ and DFT+$U$+$V$ all predict a rhombohedral distortion with an angle below $90^\circ$ (see Fig.~\ref{Fig:self_BaTiO3}(b)). When $U_\textrm{O}$ is included, only the eACBN0-based DFT+$U$+$V$ scheme retains this rhombohedral distortion (see Fig.~\ref{Fig:self_BaTiO3}(f)).

\subsection*{ZnO}

\begin{figure}[htbp]
\centering
\includegraphics[width=0.8\textwidth]{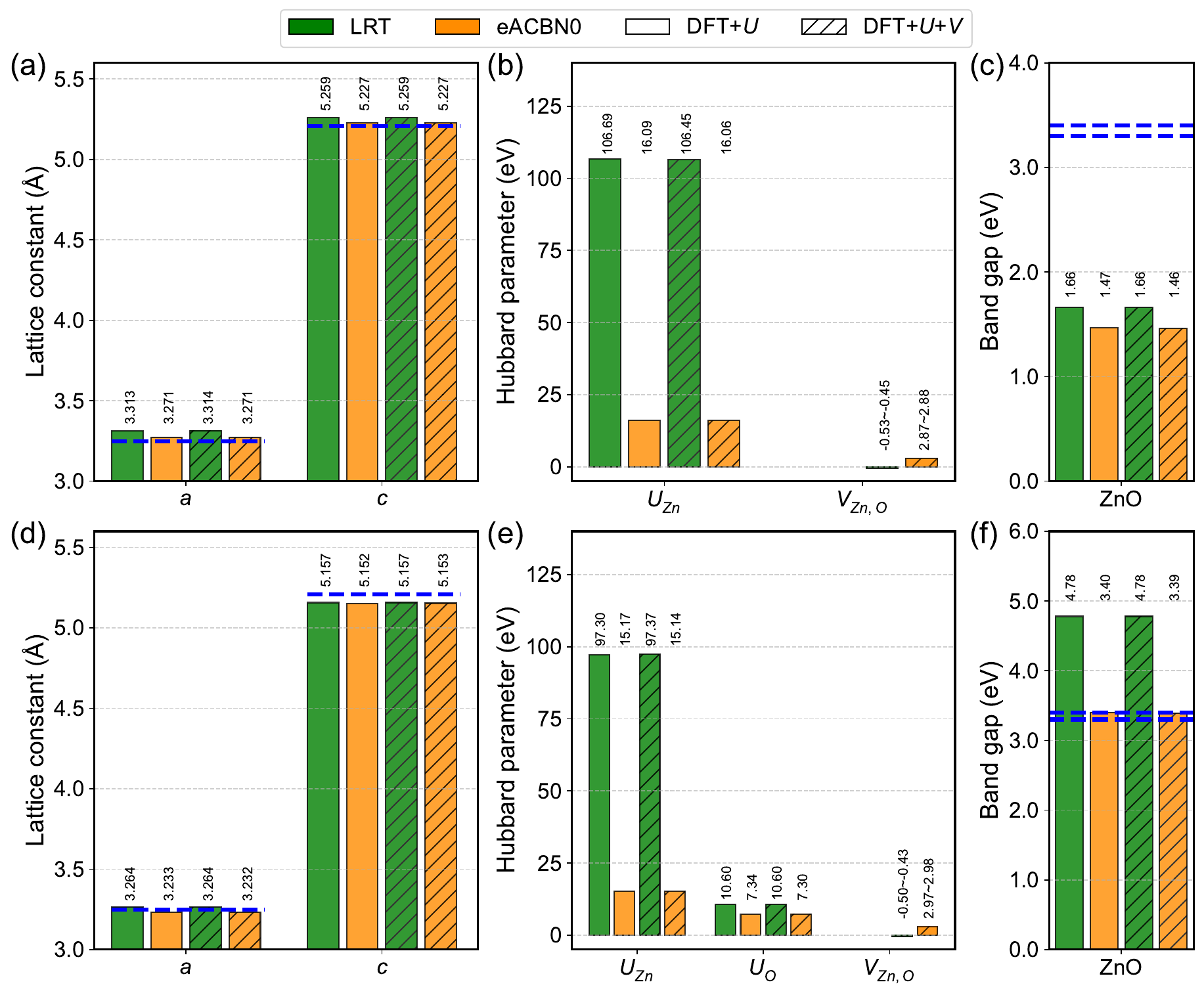}
\caption{DFT+$U$ and DFT+$U$+$V$ ``structurally self-consistent'' calculations for ZnO without $U_\textrm{O}$ [panels (a)--(c)] and with $U_\textrm{O}$ [panels (d)--(f)]. 
Panels (a) and (d) show the lattice constants $a$ and $c$. 
Panels (b) and (e) display the on-site Hubbard parameters $U_{\textrm{Zn}}$ and $U_\textrm{O}$ for Zn-$3d$ and O-$2p$ states, respectively, along with the inter-site parameter $V_{\textrm{Zn,O}}$ between Zn-$3d$ and O-$2p$ states. 
Experimental lattice constants and band gaps~\cite{Mang1995ssc,Chen:1998,Reynolds1999prb,Dong2004prb} are indicated by dashed lines in panels (a) and (d), and panels (c) and (f), respectively.
}
\label{Fig:self_ZnO}
\end{figure}

Figure~\ref{Fig:self_ZnO} presents the results for ZnO, a system with a nominally filled $d^{10}$ shell, obtained using the ``structurally self-consistent'' scheme (see Fig.~\ref{fig:self-consistent}). In line with the findings reported in the main text (without DFT+$U$+$V$ structural relaxations), the LRT approach again yields an unphysically large value of $U_{\textrm{Zn}}$. Nevertheless, this large $U_{\textrm{Zn}}$ has only a minor effect on both the lattice constants and the band gap, because the Zn-$3d$ states are fully occupied and play essentially no role in the chemical bonding or the states around the band edges.

As shown in Figs.~\ref{Fig:self_ZnO}(d)–(f), and consistent with Fig.~9 in the main text, including $U_\textrm{O}$ increases the band gap significantly. However, while LRT leads to a band gap that substantially overestimates the experimental value, the ACBN0 scheme yields a band gap in very good agreement with experiment.

\newpage
\clearpage

\section{Self-consistent Hubbard parameters using PSlibrary}

\begin{figure}[htbp]
\centering
\includegraphics[width=0.8\textwidth]{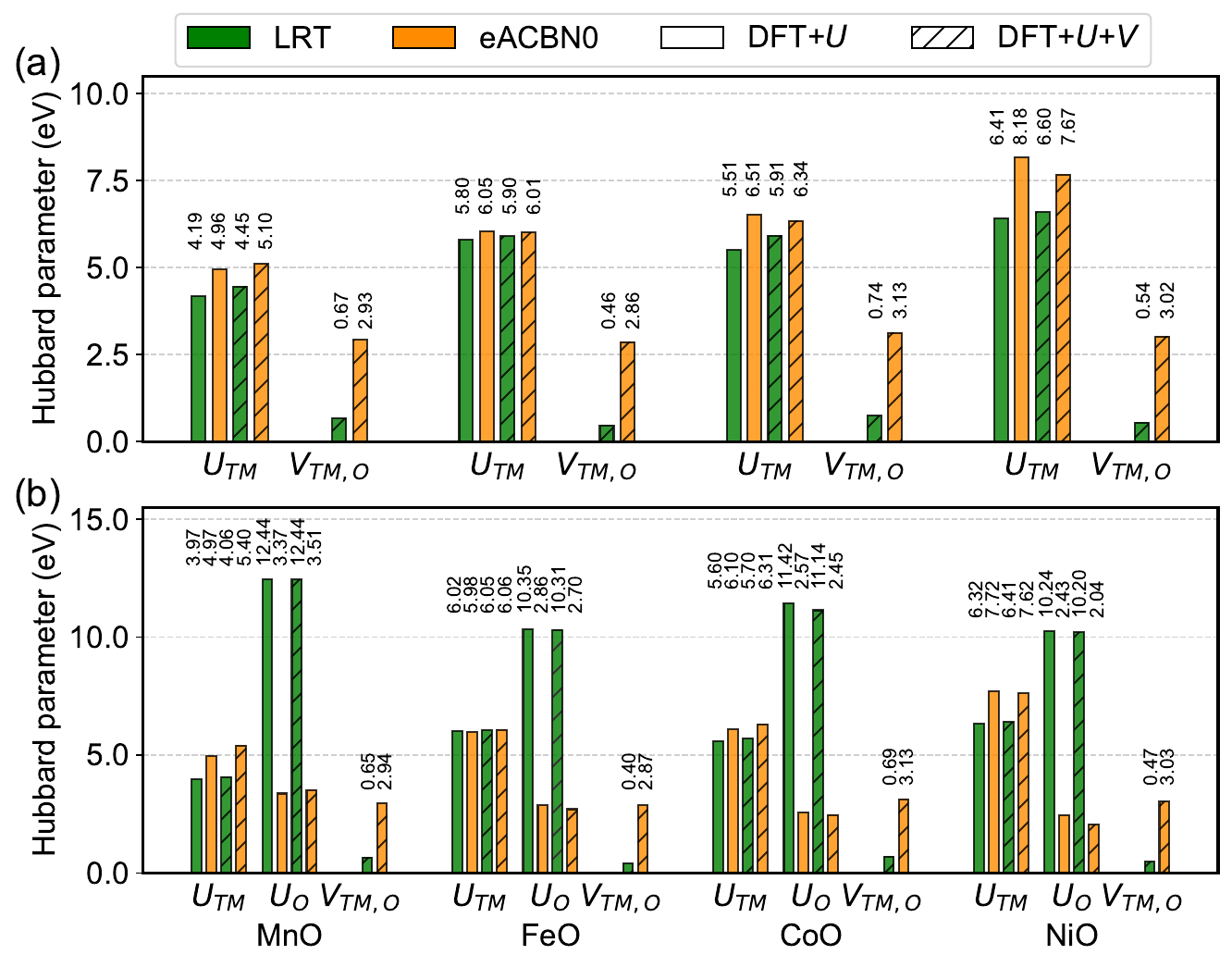}
\caption{DFT+$U$ and DFT+$U$+$V$ ``self-consistent'' calculations for TMOs using pseudopotentials from the PSlibrary (v0.3.1)~\cite{DalCorso:2014}: (a) without and (b) with the oxygen Hubbard correction ($U_\textrm{O}$). Panel (a) and (b) show the on-site Hubbard parameter $U_{\textrm{TM}}$ for the TM-$3d$ states and inter-site parameter $V_{\textrm{TM,O}}$ between TM-$3d$ and O-$2p$ states. 
LRT and eACBN0 indicate the method used to obtain the Hubbard parameters. 
}
\label{Fig:TMO_ps}
\end{figure}

Figure~\ref{Fig:TMO_ps} shows the Hubbard $U$ and $V$ parameters computed using pseudopotentials from the PSlibrary (v0.3.1)~\cite{DalCorso:2014}. The calculations were performed using the ``self-consistent'' protocol at fixed geometry. Specifically, the geometry was optimized at the PBEsol level using the same PSlibrary pseudopotentials.

\vskip 0.5cm

\end{document}